\documentclass[11pt]{article}
\usepackage[margin=1in]{geometry}
\usepackage{setspace}
\usepackage{amsmath,amssymb,amsthm}
\usepackage[utf8]{inputenc}
\usepackage[T1]{fontenc}
\usepackage{graphicx}
\usepackage{natbib}
\usepackage{hyperref}
\usepackage{url}
\usepackage{booktabs}
\usepackage{amsfonts}
\usepackage{nicefrac}
\usepackage{microtype}
\usepackage{comment}

\usepackage{xcolor}
\usepackage{graphicx}
\usepackage{subcaption}
\usepackage{makecell}

\newsavebox{\resulttable}
\newlength{\resultheight}
\usepackage{algorithm}
\usepackage{algpseudocode}

\newtheorem{theorem}{Theorem}
\newtheorem{lemma}{Lemma}
\newtheorem{proposition}{Proposition}

\newtheorem{definition}{Definition}
\newtheorem{remark}{\textit{Remark}}
\newtheorem{assumption}{Assumption}

\newcommand{\reals}{\mathbb{R}}
\newcommand{\E}{\mathbb{E}}
\newcommand{\M}{\mathcal{M}}
\newcommand{\T}{\mathcal{T}}

\newcommand{\perm}{\boldsymbol{\pi}}

\newcommand{\Ld}{L_{d}}
\newcommand{\Lg}{L_{g}}
\newcommand{\Lng}{L_{ng}}
\newcommand{\Lza}{L_{z}(\alpha)}
\newcommand{\Lda}{L_{d}(\alpha)}
\newcommand{\Lga}{L_{g}(\alpha)}
\newcommand{\Lnga}{L_{ng}(\alpha)}
\newcommand{\Da}{D^{-\alpha}}
\newcommand{\tD}{\widetilde{D}}

\newcommand{\zero}{\boldsymbol{0}}
\newcommand{\one}{\boldsymbol{1}}

\title{\textbf{Generalized Laplacian in Spectral Seriation on Manifold Data}}

\author{
    Ruizi Wu$^{*}$, 
    Wanjie Wang$^{*}$,
    Jinchi Lv$^{\dagger}$
}

\date{November 6, 2026}

\begin{document}

\maketitle

\begingroup
\renewcommand{\thefootnote}{\fnsymbol{footnote}}

\footnotetext[1]{Department of Statistics and Data Science, National University of Singapore.}

\footnotetext[2]{Data Sciences and Operations Department, University of Southern California Marshall School of Business}

\endgroup

\begin{abstract}
Noisy high-dimensional observations may have conditional means lying on a smooth closed curve, indexed by {\it temporal labels} or {\it phases}. We want to recover the latent phases for each observation and thus retrieve a circular ordering. Spectral seriation uses the graph Laplacian's spectral embedding to recover the phases and ordering. 
We consider generalized graph Laplacians in spectral seriation, where a parameter $\alpha$ controls the degree normalization and affects the accuracy and stability of both tasks. We develop and compare three generalized Laplacian constructions: a shifted generalized Laplacian, a graph-difference Laplacian, and an extra-normalized generalized Laplacian. Among these, the extra-normalized variant is less sensitive to the choice of $\alpha$ and achieves the strongest overall recovery performance. For this variant, we establish a theoretical error bound relating phase estimation accuracy to eigenspace perturbation, accounting for the inherent rotation and reflection ambiguity. 
In synthetic experiments across various scenarios, our new graph Laplacian outperforms the classic spectral seriation method and other seriation methods. On the Pink Angel data, which records photos from different viewpoints, our method achieves zero ordering error and the lowest phase recovery error. On FUCCI cell-cycle data, our method performs the best when all 6 cells are mixed, demonstrating the robustness to heterogeneity. 
\end{abstract}

\noindent\textit{Keywords:} phase recovery; spectral seriation; generalized graph Laplacians; manifold learning; spectral embedding; circular ordering.

\section{Introduction}\label{sec:intro}

Many high-dimensional datasets have an underlying one-dimensional progression. Examples include images indexed by viewing angles~\citep{whiteley2022statistical}, single-cell expression profiles along the cell cycle~\citep{Haghverdi2016}, and cryo-EM images associated with molecular motion~\citep{Moscovich2020}. Although each observation is high-dimensional, it lies on a one-dimensional manifold indexed by angles, cell cycle stage, or molecular motion stage. We call these indices {\it phases}. 
Consider the periodic settings where the possible phases form a smooth closed curve. Our goal is to estimate these latent phases of noisy observations and recover their circular ordering. These two tasks are related but distinct, and both are important for downstream scientific analysis.

To illustrate this structure, consider images of an object rotating through $360^\circ$. Each image is represented by $Z_i\in\reals^p$, and we write $Z_i=X(t_i)+e_i$, where $t_i$ is the viewing angle, $X(t_i)$ is the underlying noiseless image, and $e_i$ is observation noise. As the angle varies, the noiseless images trace a one-dimensional curve $\M=\{X(t): t\in[0,2\pi)\}$ in image space. A full rotation returns to the same image, so the curve is closed. When $X$ is smooth and regular without self-intersections, $\M$ forms a smooth closed manifold. A similar structure arises in cell-cycle data, where gene expression profiles change as cells progress through the cycle. By representing each cell's progression state with a latent phase $t_i$, we can estimate its position within the cycle from the observed gene expression profile and infer its cell-cycle stage.
 
This geometric structure suggests recovering phases by mapping the underlying curve onto a circle.
When the phase is parameterized by rescaled arc length, the first nonconstant eigenspace of the Laplace--Beltrami operator (the manifold Laplacian) on $\M$ is spanned by $\cos(t)$ and $\sin(t)$. These functions provide circular coordinates, allowing phases to be identified up to rotation and reflection. Spectral seriation uses graph-Laplacian eigenvectors to recover the ordering of observations. Building on this idea, \cite{WangSeriation} use a two-dimensional spectral embedding to estimate latent phases and recover their circular ordering. Let $S=(S_{ij})_{n\times n}$ be an affinity matrix constructed from the observations, and $D=\mathrm{diag}(\sum_{j=1}^n S_{ij}: i\in[n])$ be the degree matrix. The standard normalized graph Laplacian is $L=I-D^{-1/2}SD^{-1/2}$. Under some regularity conditions, the eigenspace spanned by its second and third eigenvectors approximates the sine--cosine eigenspace~\citep{Coifman2006,Singer2006}. The angular coordinates of this embedding yield phase estimates, which are then sorted to recover circular ordering.

The standard normalized graph Laplacian uses the degree-normalized affinity $D^{-1/2}SD^{-1/2}$. A natural generalization is to consider $S_\alpha=D^{-\alpha}SD^{-\alpha}$, where $\alpha\geq 0$ controls the strength of degree correction~\citep{LvGeneral}. The choices $\alpha=0$ and $\alpha=1/2$ recover the original affinity matrix and the standard symmetric normalization, respectively. Degree-based normalization has been studied in manifold learning and diffusion maps~\citep{Coifman2006, Singer2016}, while recent work on generalized Laplacian matrices has further highlighted the statistical role of $\alpha$ in spectral inference~\citep{LvGeneral}. Varying $\alpha$ changes how pairwise similarities are weighted by node degrees and can therefore alter the resulting spectral embedding. Since phase estimates and circular ordering are obtained from this embedding, the choice of $\alpha$ may affect the accuracy and stability of both recovery tasks.

However, the generalized affinity $S_\alpha$ alone does not determine how the graph Laplacian should be constructed. In our closed-curve setting, phase recovery relies on a two-dimensional spectral embedding whose associated eigenspace approximates $\operatorname{span}\{\cos(t),\sin(t)\}$. A direct construction such as $I-D^{-\alpha}SD^{-\alpha}$ need not preserve the spectral structure of the standard normalized Laplacian, and the desired eigenspace alignment is not guaranteed by degree correction alone. The key difficulty is thus not whether degree correction can be introduced, but how it should be incorporated into a Laplacian construction. 
This leads to a central question of this paper: \textit{which generalized Laplacian construction is suitable for latent phase recovery and circular ordering?}

To address this question, we investigate three generalized Laplacian constructions based on $S_\alpha$: a shifted generalized Laplacian, a graph-difference Laplacian, and an extra-normalized generalized Laplacian. The first two incorporate $S_\alpha$ directly into different Laplacian-type operators, whereas the third treats $S_\alpha$ as a reweighted affinity matrix and further normalizes it by its own degree matrix. This additional normalization preserves the standard normalized-Laplacian structure while allowing $\alpha$ to control the degree correction. We demonstrate that this extra-normalized construction provides a natural generalized analog of the standard normalized Laplacian for latent phase recovery and circular ordering on closed curves.

\subsection{Our contributions}

The major contributions of our paper are threefold. First, we formulate generalized spectral seriation as an operator-construction problem, rather than a direct replacement of $D^{-1/2}SD^{-1/2}$ by $D^{-\alpha}SD^{-\alpha}$. The same generalized affinity can be converted into several plausible Laplacian-type matrices, but these matrices need not preserve the eigenspace carrying the latent labels $t_i$. This identifies a structural issue absent from the standard generalized-affinity analysis and fixed-normalization spectral seriation.

Second, we propose three Laplacian-type constructions and identify the extra-normalized operator as the appropriate generalized analog of $L$ for phase recovery and circular ordering. The key novelty is the separation of two roles: $\alpha$ modifies the affinity matrix, while the final normalization restores the Laplacian geometry required. We establish convergence of the expected eigenspace to the discretized sine--cosine eigenspace and derive a conditional phase estimation error bound. This explains why the extra-normalized operator preserves the angular embedding and provides theoretical support for phase recovery.

Third, we investigate the role of the parameter $\alpha$ in phase recovery and circular ordering. Our simulations show that the extra-normalized generalized Laplacian is less sensitive to $\alpha$ than the other two constructions. Moreover, its recovery accuracy can still benefit from a suitable choice of $\alpha$, motivating us to suggest a practical value of $\alpha$. With this choice, the method achieves strong recovery performance on synthetic data, the best viewing-angle recovery on image data, and competitive cell-cycle ordering on FUCCI data.

Together, our results add theoretical and empirical depth beyond fixed-normalization spectral seriation~\citep{WangSeriation}. They unveil which generalized operator preserves the relevant two-dimensional phase eigenspace, why seemingly natural alternatives can lose the phase geometry, and how degree correction can be incorporated without sacrificing the angular recovery mechanism.

\subsection{Related works}
The seriation problem has a long history. Early work formulated seriation as a combinatorial problem~\citep{Arabie1996,Hubert2001}. Such formulations are statistically natural, but exact optimization is computationally expensive~\citep{Brusco2005,Hahsler2008}. Spectral relaxations provide a more scalable alternative through graph Laplacian eigenvectors~\citep{Atkins1998,Fogel2014}. Recent statistical works have studied convex relaxations for permutation problems~\citep{Fogel2013}, permutation recovery under monotone matrix models~\citep{Ma2021a,Ma2021b}, noisy matrix reordering~\citep{CaiMa2023}, and minimax-optimal seriation~\citep{Issartel2024}. These methods primarily focus on recovering the ordering of observations and are powerful when the affinity matrix has an ordering-compatible pattern. Beyond ordering, \cite{Giraud2023} study latent position estimation on a one-dimensional torus from pairwise affinities. More closely related to our setting, \cite{WangSeriation} use a two-dimensional graph-Laplacian embedding to recover latent phases from noisy manifold observations and then obtain their circular ordering. Building on this approach, we investigate how generalized degree normalization and different Laplacian constructions affect both phase recovery and circular ordering.

A second viewpoint treats the observations as samples from a low-dimensional manifold, so that recovering $t_i$ becomes an embedding problem. Laplacian eigenmaps and diffusion maps use graph Laplacian eigenvectors to construct nonlinear low-dimensional embeddings from local similarities~\citep{Belkin2003,Coifman2006,Belkin2008}. \cite{Patrick2020} discusses the embedding dimensions with a latent manifold structure. Isomap approximates geodesic distances by shortest paths on a neighborhood graph~\citep{Tenenbaum2000}, while t-SNE and UMAP are used widely for visualization and exploratory analysis~\citep{Maaten2008,McInnes2018}. These methods can reveal low-dimensional structure, but their embeddings do not directly yield latent phase labels. For a one-dimensional closed manifold, the trigonometric structure of the Laplace–Beltrami eigenfunctions provides a connection between spectral embeddings and latent phases.

Generalized graph Laplacians and degree-normalized affinities form a third line of related work. The family $S_\alpha=D^{-\alpha}SD^{-\alpha}$ allows the strength of degree correction to vary through $\alpha$~\citep{LvGeneral}. Initial applications of the asymptotic theory in \citep{LvGeneral} include graph neural networks and Laplacian-regularized graph smoothing. A closely related construction appears in diffusion maps, where degree-reweighted affinities are further row-normalized to form a Markov transition operator~\citep{Coifman2006}. For phase recovery and circular ordering, the relevant question, however, is not only how to define a generalized affinity, but how to convert that affinity into an operator whose leading nonconstant eigenspace approximates $\cos(t)$ and $\sin(t)$. 

\section{Problem setup and spectral seriation}\label{sec:specseriation}

Consider observations 
\begin{equation}\label{eq:dataZ}
    Z_i = X(t_i) + e_i, \qquad i \in \{1, 2, \ldots, n\},
\end{equation}
where $t_i \in[0, 2\pi)$ is the latent phase associated with observation $i$, $X:[0,2\pi] \to \reals^p$ is the underlying closed curve with $X(0)=X(2\pi)$, and $e_i \in \reals^p$ is an additive noise with $\E[e_i\mid t_i]=0$. Our goal is to recover the latent phases $t_i$ and the corresponding circular ordering $\perm:[n]\to[n]$ satisfying $t_{\perm(1)}<t_{\perm(2)}<\cdots<t_{\perm(n)}$.

Spectral seriation uses graph-Laplacian eigenvectors to recover latent orderings~\cite{Fogel2014}. For closed curves, \cite{WangSeriation} use a two-dimensional spectral embedding to estimate latent phases, recover their circular ordering, and establish theoretical guarantees under observation noise. Following this approach, we first construct an affinity matrix $S$ from the observations. Given a bandwidth parameter $h>0$, we define $S$ using the Gaussian kernel
\begin{equation}\label{eq:gaussian-kernel}
   S_{ij} 
   = k(Z_i, Z_j; h), 
   \qquad \mbox{where }
   k(x,y;h)
   =  \exp\{ - {\|x-y\|_2^2}/{2h^2} \}/\sqrt{2\pi h^2}.
\end{equation}
Denote $D = \mathrm{diag}(\sum_{j=1}^n S_{ij}: i\in[n])$ the diagonal degree matrix. The baseline method uses the standard normalized graph Laplacian
\begin{equation}\label{eq:baseline-laplacian}
    L = I - D^{-1/2} S D^{-1/2}.
\end{equation}
The spectral embedding is motivated by the eigenfunctions of the manifold Laplacian on the underlying curve. We assume that $X$ is smooth, regular, and non-self-intersecting. The latent phase $t$ is taken to be a rescaled arc-length parameter so that $\|X'(t)\|_2$ is constant. With this parameterization, the first nonconstant eigenspace of the manifold Laplacian is two-dimensional, spanned by $\cos(t)$ and $\sin(t)$. Let $F_+$ and $F_-$ be the eigenvectors associated with the second and third smallest eigenvalues of $L$. Under some regularity conditions, the eigenspace spanned by $F_+$ and $F_-$ approximates the aforementioned eigenspace. More discussion and illustrations can be found in Appendix~\ref{appsec:spectralseriation}.

Motivated by the ideal embedding $(\cos(t_i),\sin(t_i))$, we normalize each embedded point to unit length and estimate its latent phase through
\begin{equation*}
    \cos(\hat t_i)
    = \frac{F_+(i)}{\left(F_+^2(i) + F_-^2(i)\right)^{1/2}},
    \qquad
    \sin(\hat t_i)
    = \frac{F_-(i)}{\left(F_+^2(i) + F_-^2(i)\right)^{1/2}}.
\end{equation*}
Equivalently, $\hat t_i$ is the angular coordinate of $(F_+(i),F_-(i))$. Sorting the estimated phases then yields the estimated circular ordering $\hat{\pi}$. Implementation details are given in Algorithm~\ref{alg:baseline}.

\begin{algorithm}[h]
\caption{Baseline spectral seriation for closed-loop temporal recovery}
\label{alg:baseline}
\begin{algorithmic}[1]
    \Require Observations $Z_1,\dots,Z_n$; kernel bandwidth $h$.
    \Ensure Estimated phases $\hat t_1,\dots,\hat t_n\in[0,2\pi)$ and circular ordering $\hat\pi$.
    \State Construct the affinity matrix $S$ using \eqref{eq:gaussian-kernel}.
    \State Form the degree matrix $D$ and the normalized graph Laplacian $L=I-D^{-1/2}SD^{-1/2}$.
    \State Find eigenvectors $F_+$ and $F_-$ associated with the second and third smallest eigenvalues of $L$.
    \State Find $\hat{t}_i\in [0,2\pi)$ so that $\cos(\hat{t}_i)=\frac{F_+(i)}{\sqrt{(F_+(i))^2+(F_-(i))^2}}, \, \sin(\hat{t}_i)=\frac{F_-(i)}{\sqrt{(F_+(i))^2+(F_-(i))^2}}$.
    \State Sort the estimated phases to obtain $\hat\pi$ such that $\hat t_{\hat\pi(i)}$ is increasing in $i$.
    \State Return $\hat{t}=(\hat{t}_1, \hat{t}_2, \ldots, \hat{t}_n)$ and $\hat\pi$.
\end{algorithmic}
\end{algorithm}

Since the first nonconstant manifold-Laplacian eigenvalue has multiplicity two, an orthonormal basis for its eigenspace is determined only up to an orthogonal transformation $Q\in \mathbb{O}(2)$. This induces a global rotation and reflection ambiguity in the recovered phases $\hat{t}_i$. Consequently, the circular ordering is identifiable only up to a cyclic shift and reversal.

\section{Generalized Laplacian for spectral seriation}\label{sec:generalLap}

The baseline spectral seriation method uses the normalized graph Laplacian $L=I-D^{-1/2}SD^{-1/2}$. Following the generalized degree normalization~\cite{LvGeneral}, we consider the generalized affinity matrix
\begin{equation}\label{eq:L_z}
    S_\alpha = D^{-\alpha}SD^{-\alpha}, \qquad \alpha\geq 0.
\end{equation}
The parameter $\alpha$ controls the strength of degree correction: $S_0=S$, while $S_{1/2}=D^{-1/2}SD^{-1/2}$. However, the generalized affinity alone does not specify a Laplacian operator, and different constructions can lead to different spectral embeddings.

For spectral seriation, the object of interest is a Laplacian-type operator whose two-dimensional nontrivial eigenspace can approximate $\operatorname{span}\{\cos(t),\sin(t)\}$. We therefore investigate how the same generalized affinity $S_\alpha$ can be incorporated into different Laplacian constructions for phase and circular ordering recovery.

\subsection{Three Laplacian variants}\label{subsec:three-variants}

We consider three Laplacian variants derived from $S_\alpha$. Each represents a different way to construct a Laplacian-type operator from the generalized affinity.

{\bf Shifted generalized Laplacian.} Our first variant shifts the generalized affinity to obtain a positive semidefinite matrix:
\begin{equation}\label{eq:Ld}
    \Lda 
    = d_{\max}(\alpha) I - S_\alpha,
\end{equation}
where $d_{\max}(\alpha) = \max_{i\in[n]} D_{ii}^{\,1-2\alpha}$. Indeed, $\Lda=[d_{\max}(\alpha)I-D^{1-2\alpha}]+D^{-\alpha}(D-S)D^{-\alpha}$ is the sum of two positive semidefinite matrices. This variant preserves the eigenspaces of the direct construction $I-S_\alpha$ while ensuring positive semidefiniteness. When $\alpha=1/2$, it coincides with the baseline Laplacian $L$.

{\bf Graph-difference Laplacian.} Our second variant preserves the intuition that a Laplacian should encode the difference between node degree and pairwise affinity:
\begin{equation}\label{eq:Lg}
    \Lga 
    = D^{-\alpha}(D-S)D^{-\alpha}
    = D^{1-2\alpha}-S_\alpha.
\end{equation}
It reduces to the unnormalized graph Laplacian at $\alpha=0$, and to the baseline Laplacian at $\alpha=1/2$.

{\bf Extra-normalized generalized Laplacian.} Our third variant is motivated by~\cite{Singer2016}. We treat $S_\alpha$ as a reweighted affinity matrix and normalize it using its own degree matrix, $\widetilde D_\alpha=\mathrm{diag}(S_\alpha\one)$. Define
\begin{equation}\label{eq:Lng}
    \Lnga 
    = I -\tD_\alpha^{-1/2} S_\alpha \tD_\alpha^{-1/2}.
\end{equation}
This construction retains the symmetric normalized-Laplacian form, with $L_{ng}(0)=L$.

The three variants use the same generalized affinity $S_\alpha$, but differ in how this affinity enters the final operator. These differences can affect the nontrivial eigenspace used for phase recovery, especially when node degrees vary. 

To illustrate the role of extra normalization, consider a simplified case $D=dI$, where $d>0$. Then $S_\alpha=d^{-2\alpha}S$ and $\tD_\alpha=d^{1-2\alpha}I$. Consequently,
\begin{equation*}
    \Lda = \Lga = d^{1 - 2\alpha}L,
    \qquad 
    \Lnga = L.
\end{equation*}
Thus, all three variants share the same eigenspaces in this simplified case. Extra normalization also removes the degree scaling introduced by $\alpha$. and recovers the baseline Laplacian $L$. When node degrees are heterogeneous, however, the three variants may have different eigenspaces. Section~\ref{sec:theory} studies the extra-normalized construction under our sampling and noise assumptions.

\begin{remark}
Diffusion maps~\citep{Coifman2006} constructs low-dimensional embeddings from the transition matrix $P_\alpha=\tD_\alpha^{-1}\Da S\Da$. Thus, our extra-normalized generalized Laplacian is the symmetric normalized Laplacian associated with this diffusion operator. Our focus is on how $\alpha$ affects phase and circular ordering recovery under different Laplacian generalizations.
\end{remark}

\subsection{Generalized spectral seriation}

We now describe a unified phase-recovery pipeline for the three variants. For a fixed $\alpha$ and $\star\in\{d,g,ng\}$, let $F_{\star,+},F_{\star,-}\in\reals^n$ be eigenvectors associated with the second and third smallest eigenvalues of $L_\star(\alpha)$. As in the baseline method, we recover $\hat{t}$ from the angular coordinates:
\begin{equation*}
    \cos(\hat{t}_i)
    = F_{\star,+} \left( F_{\star,+}^2(i) + F_{\star,-}^2(i) \right)^{-\frac{1}{2}},
    \qquad
    \sin(\hat{t}_i)
    =F_{\star,-}(i) \left( F_{\star,+}^2(i) + F_{\star,-}^2(i) \right)^{-\frac{1}{2}}.
\end{equation*}
Sorting $\hat{t}$ yields the recovered circular ordering $\hat{\pi}$. Algorithm~\ref{alg:generalized} summarizes the procedure.

\begin{algorithm}[htb]
\caption{Temporal label recovery via generalized Laplacians}\label{alg:generalized}
\begin{algorithmic}[1]
    \Require Observations $Z_1,\ldots,Z_n$; kernel bandwidth $h$; parameter $\alpha$; method $\star\in\{d,g,ng\}$.
    \Ensure Estimated phases $\hat{t}=(\hat{t}_1, \hat{t}_2, \ldots, \hat{t}_n) \in[0,2\pi)$ and circular ordering $\hat{\pi}$.
    \State Construct $S$ using \eqref{eq:gaussian-kernel}, its degree matrix $D$, and $S_\alpha=D^{-\alpha}SD^{-\alpha}$.
    \If{$\star=d$}
        \State Set $L_\star(\alpha)=d_{\max}(\alpha)I-S_\alpha$.
    \ElsIf{$\star=g$}
        \State Set $L_\star(\alpha)=D^{-\alpha}(D-S)D^{-\alpha}$.
    \ElsIf{$\star=ng$}
        \State Set $\tD_\alpha=\mathrm{diag}(S_\alpha\one)$ and $L_\star(\alpha)=I-\tD_\alpha^{-1/2}S_\alpha\tD_\alpha^{-1/2}$.
    \EndIf
    \State Find eigenvectors $F_+$ and $F_-$ associated with the second and third smallest eigenvalues of $L_\star(\alpha)$.
    \State Find $\hat{t}_i\in [0,2\pi)$ so that
    $\cos(\hat{t}_i)=\frac{F_+(i)}{\sqrt{(F_+(i))^2+(F_-(i))^2}}, \, \sin(\hat{t}_i)=\frac{F_-(i)}{\sqrt{(F_+(i))^2+(F_-(i))^2}}.$
    \State Return $\hat{t}=(\hat{t}_1, \hat{t}_2, \ldots, \hat{t}_n)$ and $\hat\pi$.
\end{algorithmic}
\end{algorithm}

Although Algorithm~\ref{alg:generalized} applies to all three variants, our theoretical analysis focuses on the extra-normalized generalized Laplacian $\Lnga$. Section~\ref{sec:theory} studies the convergence of the expected eigenspace and derives a conditional phase estimation error bound for the extra-normalized generalized Laplacian.

\section{Theoretical guarantees}\label{sec:theory}

To study the convergence of the phase estimator $\hat{t}$ obtained by Algorithm~\ref{alg:generalized} with $\Lng$, we relate its estimation error to the convergence of the two-dimensional eigenspace toward the target cosine--sine space. Instead of analyzing the observed Laplacian directly, we decompose the affinity matrix $S$ into its conditional expectation and a residual noise component. We first establish eigenspace convergence for the Laplacian constructed from the conditional expectation, and then derive a conditional phase estimation error bound in terms of the observed-versus-expected Laplacian perturbation.

\subsection{Definitions}

Recall $Z_i=X(t_i)+e_i$ from \eqref{eq:dataZ} and the Gaussian affinity matrix $S\in\reals^{n\times n}$ \eqref{eq:gaussian-kernel}. For the theory, we assume $t_i\stackrel{\mathrm{i.i.d.}}{\sim}\mathrm{Unif}(0,2\pi)$ and independent Gaussian noise $e_i\sim N(0,\sigma^2 I_p)$, together with the curve and bandwidth conditions in Appendix~\ref{appsec:assump}. We decompose $S=H+W$, where $H=\mathbb{E}[S\mid t]$ and $W=S-H$. The entries of $H$ satisfy: for $i=j$, $H_{ij} = (2\pi h^2)^{-1/2}$; and for $i\ne j$,
\begin{equation*}
    H_{ij} = \frac{1}{\sqrt{2\pi h^2}} \left( \frac{h^2}{h^2 + 2\sigma^2} \right)^\frac{p}{2} \exp\left\{ - \frac{\|X(t_i) - X(t_j)\|_2^2}{2(h^2 + 2\sigma^2)} \right\}.
\end{equation*}
For simplicity, denote $\tau:=h^2+2\sigma^2$ and $C_{h,\tau}:=h^p\tau^{-p/2}/\sqrt{2\pi h^2}$. The diagonal entries can also be written as $H_{ii}=C_{h,\tau}(1+\rho_{h,\tau,p})$, where $\rho_{h,\tau,p}:=\tau^{p/2}h^{-p}-1$ measures their discrepancy from the Gaussian form of the off-diagonal entries. This discrepancy contributes an additional term to the eigenspace error.

\begin{definition}\label{def:LMatrix}
    For any symmetric nonnegative matrix $A\in\reals^{n\times n}$ with positive degrees, let $D_A = \mathrm{diag}(A\one)$ be the diagonal degree matrix. The $\alpha$-normalized matrix is $\widetilde{A}_{\alpha}:=D_A^{-\alpha}AD_A^{-\alpha}$, with degree matrix $\tD_{\alpha,A} = \mathrm{diag}(\widetilde{A}_{\alpha}\one)$. Define the extra-normalized Laplacian $L^\star_{\alpha, A} := I - \tD_{\alpha, A}^{-1/2} \widetilde{A}_{\alpha} \tD_{\alpha, A}^{-1/2}$, and let $F_{\alpha,A} \in\reals^{n\times 2}$ have orthonormal columns spanning the eigenspace associated with its second and third smallest eigenvalues.
\end{definition}

As mentioned in Section \ref{sec:specseriation}, $t$ is identifiable only up to rotation and reflection. To account for such ambiguity, we define the timestamp estimation error of $\hat{t}$ as
\begin{equation}\label{eq:estError}
    \mathrm{Error}(\hat{t}, t) := \min_{r\in\{-1, 1\}, \theta\in[0, 2\pi]} \frac{1}{\sqrt{n}} \left\| [rt+\theta - \hat{t}\,]_{2\pi} \right\|_2,
\end{equation}
where $[a]_{2\pi}$ denotes the representative of $a$ modulo $2\pi$ in $[-\pi,\pi]$, $r$ accounts for reflection, and $\theta$ represents a global angular shift.

\subsection{Main results}

We state our main results under Assumptions~\ref{aspt:setup}--\ref{aspt:egap}, given in Appendix~\ref{appsec:assump}. Here $c_{\mathrm{gap}}>0$ denotes the eigengap constant, with the target eigenspace separated from the remaining spectrum by at least $c_{\mathrm{gap}}\tau$. We first establish convergence of the expected eigenspace to the sampled trigonometric eigenspace.

\begin{proposition}[Expected eigenspace convergence]\label{Main_prop}
Assume that Assumptions \ref{aspt:setup}--\ref{aspt:egap} hold. Define the rows of $\Psi_n\in\reals^{n\times 2}$ as $(\Psi_n)_i = (\cos(t_i), \, \sin(t_i))$, and denote by $\Phi_n = \Psi_n (\Psi_n^\top \Psi_n)^{-1/2}$. Let $\alpha\in[0,1]$. Then we have that for all sufficiently large $n$, with probability $1 - O(n^{-2})$,
    \begin{equation}
    \label{eq:ConvH}
        \left\| F_{\alpha,H}F_{\alpha,H}^{\top} - \Phi_n\Phi_n^{\top} \right\|_2
        \le C(1 + \alpha) \left( \tau + \frac{\sqrt{\log{n}}}{n^{1/2}\tau^{5/4}} + \frac{1}{c_{\mathrm{gap}}} \frac{\rho_{h,\tau,p}}{n\tau^{3/2}} \right).
    \end{equation}
    Equivalently, there exists some $Q\in \mathbb{O}(2)$ such that with probability $1 - O(n^{-2})$,
    \begin{equation}
        \|F_{\alpha,H}Q - \Phi_n\|_F
        \le C(1 + \alpha) \left( \tau + \frac{\sqrt{\log{n}}}{n^{1/2}\tau^{5/4}} + \frac{1}{c_{\mathrm{gap}}} \frac{\rho_{h,\tau,p}}{n\tau^{3/2}} \right).
    \end{equation}
\end{proposition}

The proof of Proposition~\ref{Main_prop} is provided in Appendix~\ref{appsec:prop}. Under Assumption~\ref{aspt:egap}, the error bound vanishes as $n \to \infty$. This establishes convergence of the expected eigenspace to the sampled trigonometric eigenspace up to an orthogonal transformation.

\begin{theorem}[Estimator error bound]\label{Main_theo}
Assume that Assumption \ref{aspt:setup}--\ref{aspt:egap} hold. Let $\alpha\in[0,1]$, and assume further that $\|L^\star_{\alpha,S}-L^\star_{\alpha,H}\|_2\le U_S$ for some nonnegative number $U_S$. Then we have that with probability $1-O(n^{-2})$,
    \begin{equation}
    \label{eq:theo_error}
        \mathrm{Error}(\hat{t}, t)
        \le C \left[ (1 + \alpha) \left( \tau + \frac{\log(n)^{1/2}}{n^{1/2} \tau^{5/4}} + \frac{1}{c_{\mathrm{gap}}} \frac{\rho_{h,\tau,p}}{n\tau^{3/2}} \right) + \frac{U_S}{c_{\mathrm{gap}}\tau} \right].
    \end{equation}
\end{theorem}

The proof of Theorem 1 is provided in Appendix~\ref{appsec:theo}. Theorem~\ref{Main_theo} separates the expected eigenspace error in Proposition~\ref{Main_prop} from the observed perturbation term $U_S/(c_{\mathrm{gap}}\tau)$. Since the target two-dimensional eigenspace has a gap of order $\tau$, Davis--Kahan converts the operator perturbation into an eigenspace bound. Its eigenvectors cannot be compared individually: an orthonormal basis may rotate or reflect. We therefore use Procrustes alignment before converting the embedding error into circular phase error. The difference between $\Phi_n$ and the scaled circular embedding $\sqrt{2/n}\Psi_n$ is absorbed into the finite-sample term. In particular, the bound implies phase consistency if the errors in Proposition~\ref{Main_prop} vanish and $U_S=o(\tau)$, provided the perturbation event holds with probability tending to one.

\begin{remark}
Theorem~\ref{Main_theo} is conditional on $\|L^\star_{\alpha,S}-L^\star_{\alpha,H}\|_2 \le U_S$, whose probability is not bounded here. If $\mathbb{P}(\|L^\star_{\alpha,S}-L^\star_{\alpha,H}\|_2 \le U_S)\ge1-\delta_n$, the phase-error bound holds with probability at least $1-O(n^{-2})-\delta_n$.
Establishing such a bound requires addressing two challenges: affinity entries sharing an observation are dependent, even conditional on the latent phases; and their fluctuations propagate through both stages of degree normalization. This requires joint control of the affinity and degree perturbations. Such an analysis is beyond the scope of the present paper and will be pursued in subsequent work.
\end{remark}

\begin{remark}
When the diagonal correction and observed perturbation terms are negligible compared with the remaining terms, the error bound is dominated by $\tau + \frac{\sqrt{\log(n)}}{\sqrt{n}\tau^{5/4}}$. Balancing these terms gives $\tau \asymp (\log(n)/n)^{2/9}$ and a corresponding conditional error bound of order $(\log(n)/n)^{2/9}$, provided the theorem assumptions hold.
\end{remark}

\section{Numerical experiments}\label{sec:simulation}

We consider a synthetic closed-loop manifold in $\reals^p$. The underlying manifold is defined as $X(t) = Qv(t) \in \reals^p$, where $v(t) \in \mathbb{R}^r$ is a low-dimensional vector and $Q \in \mathbb{R}^{p \times r}$ embeds it into $p$-dimensional space with $Q^\top Q = I_r$. We generate the data points as 
\begin{equation*}
    Z_i 
    = Qv(t_i) + s(t_i)e_i, 
    \quad t_i \sim \mathrm{Unif}(0, 2\pi), 
    \quad 1 \leq i \leq n,
\end{equation*}
where $e_i \sim \mathcal{N}(\zero, \sigma^2 I_p)$ and $s(t)$ introduces heteroscedastic noise. Additional experiments with non-uniformly sampled latent phases are provided in Appendix~\ref{appsec:nonUnif}.

We use two metrics to evaluate recovery. The timestamp estimation error $\mathrm{Error}(\hat{t},t)$ measures the $\ell_2$ discrepancy between the true and estimated timestamps under optimal rotational or reflection-rotational alignment, as defined in \eqref{eq:estError}. We also introduce an ordering error to assess recovery of the circular ordering, enabling comparisons with methods that do not directly estimate timestamps:
\begin{equation}
    \mathrm{Error}(\hat{\pi},\pi)
    = \min_{\substack{r\in\{\pm1\}\\ k\in\{0,\ldots,N-1\}}}  \left[ \frac{1}{N}\sum_{i=1}^{N} \left( \frac{d_N(\hat{\pi}(i)+k,r\pi(i))}{N} \right)^2 \right]^{1/2},
\end{equation}
where $d_N(a,b):=\min_{j\in\mathbb{Z}}|a-b+jN|$.

\subsection{The role of $\alpha$ in three variants}\label{subsec:experiment1}
\begin{figure}[h]
    \centering
    \includegraphics[width=0.8\linewidth]{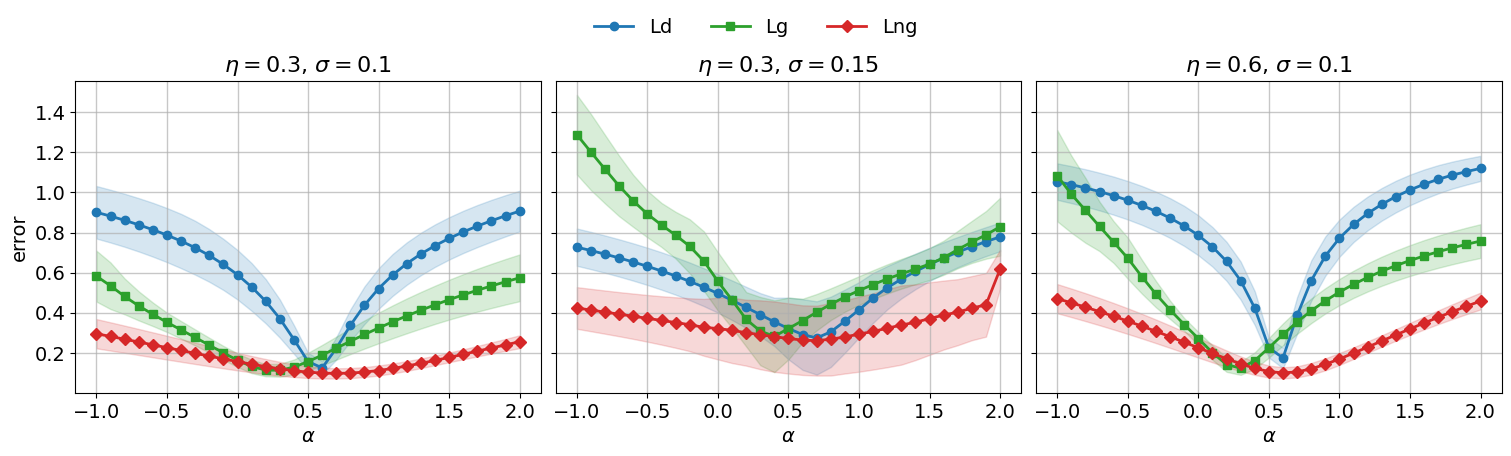}
    \caption{Mean estimation error versus $\alpha$ for three generalized Laplacians: $\Ld$ (blue), $\Lg$ (green), and $\Lng$ (red). The shaded region represents the standard deviation over 100 repetitions.}
    \label{figN2026:Experiment1_Error}
\end{figure}

We compare our three Laplacian variants $\Lda$, $\Lga$, and $\Lnga$, and examine their sensitivity to $\alpha$. 
Fix sample size $n=500$ and dimensionality $p=40$. Let $v(t)=(1.5\cos(t), \sin(t), \cos(2t), \sin(2t))^\top$ and $s(t) = 1 + \eta \cos(t - \pi/3)$. We consider three settings: a) baseline: $\eta=0.3$, $\sigma=0.10$; b) large noise with baseline heterogeneity: $\eta=0.3$, $\sigma=0.15$; c) baseline noise with large heterogeneity: $\eta=0.6$, $\sigma=0.10$. For each setting, we vary $\alpha \in \{-1, -0.9, \ldots, 2\}$, including negative values for readers who may be interested in this regime.
Figure \ref{figN2026:Experiment1_Error} reports the mean estimation error with standard deviation among 100 repetitions.

As shown in Figure~\ref{figN2026:Experiment1_Error}, the choice of $\alpha$ substantially affects the estimation error of all three methods. While $\Ld$ exhibits a sharp minimum and its error increases rapidly as $\alpha$ deviates from the optimum, $\Lng$ maintains relatively low estimation error over a much broader range of $\alpha$. This advantage is particularly evident under larger heterogeneity, where the other two variants become more sensitive to the choice of $\alpha$. Although $\Lng$ does not always achieve the lowest error at every $\alpha$, it generally performs well even when $\alpha$ is not optimally selected. These results demonstrate that extra normalization improves robustness to the choice of $\alpha$, reducing the need for precise parameter tuning. Additional comparisons across sample sizes, bandwidths, and similarity constructions are provided in Appendices~\ref{appsubsec:sampleSize}--\ref{appsubsec:DiffSimConst}.

We further examine the empirically optimal $\alpha$ for each Laplacian variant. Fix the heterogeneity parameter $\eta=0.6$ and take the noise level $\sigma=\{0.050, 0.075, \ldots, 0.150\}$. For each $\sigma$, the optimal $\alpha$ is recorded for each repetition. Table \ref{tabN2026:Experiment1_alpha} reports the median and interquartile range of these optimal $\alpha$ values among 100 repetitions for each $\sigma$.

\begin{table}[h]
\caption{Median optimal $\alpha$ with ($25\%$ quantile, $75\%$ quantile).}
\label{tabN2026:Experiment1_alpha}
\centering
\scriptsize
\begin{tabular}{lcccccc}
\toprule
Fix $\eta=.60$
& $\sigma = .050$ & $\sigma = .075$ & $\sigma = .100$ & $\sigma = .125$ & $\sigma = .150$ \\
\midrule
$\Lda$
& \makecell{.500 \space (.500, .500)}
& \makecell{.500 \space (.500, .500)}
& \makecell{.600 \space (.600, .600)}
& \makecell{.600 \space (.600, .600)}
& \makecell{.700 \space (.600, .700)} \\
\addlinespace[3pt]
$\Lga$
& \makecell{.200 \space (.200, .300)}
& \makecell{.200 \space (.200, .300)}
& \makecell{.300 \space (.200, .300)}
& \makecell{.300 \space (.300, .400)}
& \makecell{.400 \space (.300, .400)} \\
\addlinespace[3pt]
$\Lnga$
& \makecell{.600 \space (.500, .600)}
& \makecell{.600 \space (.500, .600)}
& \makecell{.600 \space (.500, .700)}
& \makecell{.600 \space (.500, .725)}
& \makecell{.650 \space (.500, .800)} \\
\bottomrule
\end{tabular}
\end{table}

Table~\ref{tabN2026:Experiment1_alpha} shows that the optimal $\alpha$ for $\Ld$ and $\Lg$ is tightly concentrated within each noise setting, consistent with their greater sensitivity to $\alpha$ observed in Figure~\ref{figN2026:Experiment1_Error}. In contrast, $\Lng$ exhibits a wider spread of optimal $\alpha$ values while maintaining low estimation error over a broad range, demonstrating greater robustness to parameter tuning. Moreover, its median optimal $\alpha$ remains close to 0.6 across noise levels, motivating $\alpha=0.6$ as an empirical default. Results over broader ranges of $\eta$ and $\alpha$ further support this choice (Appendix~\ref{appsubsec:exp1}).

\subsection{Comparison with existing methods}\label{subsec:experiment2}

We compare the proposed $L_{ng}(0.6)$ with the spectral-seriation baseline $L_d(0.5)$~\citep{WangSeriation} and diffusion maps with $\alpha=0.5$~\citep{Singer2016}. For diffusion maps, we estimate latent phases from the angular coordinates of its two-dimensional spectral embedding and obtain the ordering by sorting the estimated phases. We also consider three manifold learning methods: t-SNE~\citep{Maaten2008}, UMAP~\citep{McInnes2018}, and Isomap~\citep{Tenenbaum2000}. However, their embeddings do not directly represent latent phases. By taking a one-dimensional embedding, we can obtain an estimated ordering by sorting its coordinates. Therefore, we evaluate these three methods only in terms of ordering error, while the spectral methods are evaluated using both ordering and timestamp estimation errors. Implementation details and tuning parameters are provided in Appendix~\ref{appsubsec:exp2}.

\begin{table}[b]
\centering
\caption{Mean ordering and timestamp estimation errors (with standard deviation).}
\label{tabN2026:Comparison}
\scriptsize
\begin{tabular}{c|*{6}{c}|*{3}{c}}
\toprule
& \multicolumn{6}{c|}{Ordering Error}
& \multicolumn{3}{c}{Timestamp Estimation Error} \\
\midrule
$\sigma$
& t-SNE & UMAP & Isomap & Baseline & DiffMap
& $\Lng(0.6)$
& Baseline & DiffMap & $\Lng(0.6)$ \\
\midrule
$.05$
& \makecell{.1762\\(.0294)}
& \makecell{.1260\\(.0664)}
& \makecell{.1965\\(.0058)}
& \makecell{\textbf{.0091}\\(.0005)}
& \makecell{\textbf{.0091}\\(.0005)}
& \makecell{\textbf{.0091}\\(.0005)}
& \makecell{.0890\\(.0193)}
& \makecell{.0619\\(.0060)}
& \makecell{\textbf{.0608}\\(.0056)} \\
\addlinespace[3pt]
$.10$
& \makecell{.1823\\(.0181)}
& \makecell{.0708\\(.0652)}
& \makecell{.1893\\(.0071)}
& \makecell{\textbf{.0173}\\(.0008)}
& \makecell{\textbf{.0173}\\(.0008)}
& \makecell{\textbf{.0173}\\(.0008)}
& \makecell{.1229\\(.0125)}
& \makecell{.1095\\(.0055)}
& \makecell{\textbf{.1084}\\(.0050)} \\
\addlinespace[3pt]
$.15$
& \makecell{.1581\\(.0344)}
& \makecell{.0704\\(.0412)}
& \makecell{.1739\\(.0076)}
& \makecell{\textbf{.0257}\\(.0012)}
& \makecell{\textbf{.0257}\\(.0012)}
& \makecell{\textbf{.0257}\\(.0012)}
& \makecell{.1677\\(.0088)}
& \makecell{.1601\\(.0059)}
& \makecell{\textbf{.1596}\\(.0060)} \\
\addlinespace[3pt]
$.20$
& \makecell{.1391\\(.0335)}
& \makecell{.1121\\(.0182)}
& \makecell{.1741\\(.0072)}
& \makecell{\textbf{.0345}\\(.0016)}
& \makecell{\textbf{.0345}\\(.0017)}
& \makecell{\textbf{.0345}\\(.0017)}
& \makecell{.2179\\(.0083)}
& \makecell{\textbf{.2138}\\(.0083)}
& \makecell{\textbf{.2138}\\(.0087)} \\
\bottomrule
\end{tabular}
\end{table}

We set sample size $n=500$, dimensionality $p=100$. Let $v(t)=(\cos t,\sin t)^\top$ and $s(t)=1+0.3\cos t$. We consider $\sigma\in\{0.05,0.10,0.15,0.20\}$ and report means and standard deviations over 100 repetitions. Table~\ref{tabN2026:Comparison} presents ordering errors for all six methods and timestamp estimation errors for the three spectral methods.

The left panel of Table~\ref{tabN2026:Comparison} shows that the three spectral seriation methods achieve substantially lower ordering errors than t-SNE, UMAP, and Isomap at every noise level. Our method $\Lng(0.6)$ closely matches the best ordering performance, with only small differences among the three spectral methods. For timestamp recovery, the right panel on Table~\ref{tabN2026:Comparison} shows that $\Lng(0.6)$ achieves the lowest mean error at every noise level. Its advantage over DM is modest, while the improvement over Baseline is more pronounced at lower noise levels. Together, these results show that $\Lng(0.6)$ combines near-best ordering recovery with the best timestamp estimation performance among the compared methods.

\section{Real Data}\label{sec:real}

\subsection{Pink Angel data recovery}

We further apply the phase recovery methods to the Pink Angel photo dataset (\url{https://aloi.science.uva.nl/}). This dataset contains $n=72$ images observed from equally spaced viewing angles $t_i = {2\pi(i-1)}/{72}$, $i=\{1, \ldots, 72\}$. Our goal is to recover both the circular ordering and the viewing angle of each image. Preprocessing details and the full set of images are provided in Appendix~\ref{appsubsec:angel}. All methods use the same preprocessed feature matrix, and we fix $\alpha=0.6$ for $L_{ng}$ and $\alpha=0.5$ for both Baseline and Diffusion Map.

\begin{figure}[h]
\centering

\sbox{\resulttable}{%
    \setlength{\tabcolsep}{4pt}
    \scriptsize
    \begin{tabular}{lcc}
        \toprule
        Method & Ordering Error & Timestamp Error \\
        \midrule
        t-SNE & .1857 & / \\
        \addlinespace[3pt]
        UMAP & .1873 & / \\
        \addlinespace[3pt]
        Isomap & .1927 & / \\
        \addlinespace[3pt]
        Baseline & \textbf{0.0} & .1339 \\
        \addlinespace[3pt]
        Diffusion Map & \textbf{0.0} & .1042 \\
        \addlinespace[3pt]
        $\Lng(0.6)$ & \textbf{0.0} & \textbf{.1004} \\
        \bottomrule
    \end{tabular}%
}

\setlength{\resultheight}{%
    \dimexpr\ht\resulttable+\dp\resulttable\relax
}

\begin{minipage}[t]{0.48\linewidth}
    \centering
    \vspace{0pt}
    \includegraphics[
        height=\resultheight
    ]{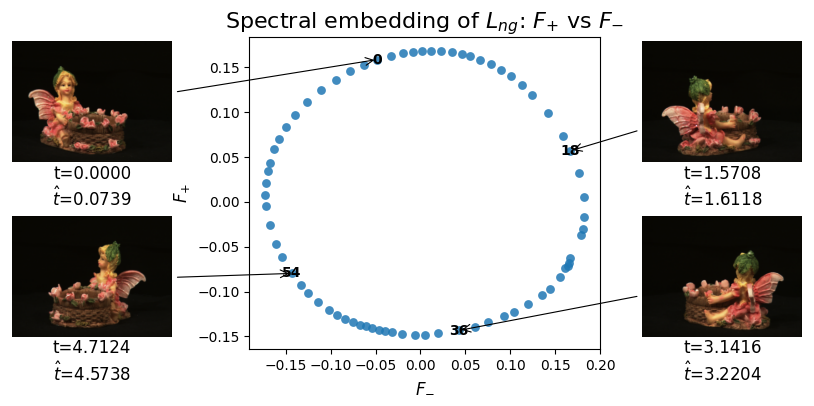}
    \par
    \subcaption{Recovered embedding}
    \label{fig:embedding}
\end{minipage}%
\hfill
\begin{minipage}[t]{0.48\linewidth}
    \centering
    \vspace{0pt}
    \usebox{\resulttable}
    \par
    \subcaption{Estimation error}
    \label{fig:estimation_error}
\end{minipage}

\caption{Phase recovery on image data. (a) Two-dimensional spectral embedding obtained by $L_{ng}(0.6)$, with representative images labeled by their true and estimated viewing angles. (b) Ordering and timestamp estimation errors for all methods.}
\label{figN2026:Real_Angel}
\end{figure}

The leading eigenvectors $F_{+}$ and $F_{-}$ are visualized in Figure~\ref{figN2026:Real_Angel}(a). The two-dimensional embedding $(F_{-},F_{+})$ forms an approximately circular pattern, consistent with the expected structure $(\cos(t),\sin(t))$. Images at four representative angles show that the estimated angle $\hat{t}$ matches closely the true angle $t$, indicating that our $\Lng$ recovers the circular ordering successfully. In the right panel of Figure~\ref{figN2026:Real_Angel}, we further compare the ordering and angle recovery error. All three spectral seriation methods recover the circular ordering exactly. For angle estimation, $\Lng(0.6)$ achieves the lowest error. These results show that our method recovers the correct image ordering while providing the most accurate viewing-angle estimates among the compared methods. Additional results on image-angle recovery using tangent distance are provided in Appendix~\ref{appsubsec:angle_TD}.

\subsection{FUCCI cell-cyle Recovery}
We evaluate our method on the FUCCI cell-cycle dataset (GSE121265; \cite{FUCCI_cite}), using gene expression profiles from 888 human induced pluripotent stem cells across six cell lines. FUCCI-derived phases serve as references for evaluating the recovered cyclic order. We analyze each cell line separately and all six jointly. Preprocessing details are provided in Appendix~\ref{appsubsec:FUCCI}. We fix $\alpha=0.6$ for $L_{ng}$ and $\alpha=0.5$ for both Baseline and Diffusion Map.

As shown in Table ~\ref{tabN2026:fucci}, spectral seriation methods achieve the lowest circular rank MSE in all six individual cell lines. Our $\Lng(0.6)$ achieves the lowest error on NA19101 and ties with diffusion maps on NA18855 at the reported precision, while the baseline performs best on the remaining four lines. On the pooled dataset, $\Lng(0.6)$ achieves the lowest error among all compared methods, highlighting its advantage in handling heterogeneous data.

\begin{table}[h]
\centering
\caption{ordering error on the FUCCI dataset.
The lowest error in each column is shown in bold.}
\label{tabN2026:fucci}
\scriptsize
\begin{tabular}{lccccccc}
\toprule
Method
& NA18511 & NA18855 & NA18870
& NA19098 & NA19101 & NA19160 & All Data \\
\midrule
t-SNE
& .2602 & .2644 & .2569
& .2500 & .2503 & .2560 & .2584 \\
\addlinespace[3pt]
UMAP
& .2572 & .2658 & .2623
& .2378 & .2575 & .2637 & .2676 \\
\addlinespace[3pt]
Isomap
& .2580 & .2632 & .2629
& .2284 & .2615 & .2606 & .2649 \\
\addlinespace[3pt]
Baseline
& \textbf{.2416} & .2508 & \textbf{.2458}
& \textbf{.2270} & .2495 & \textbf{.2424} & .2592 \\
\addlinespace[3pt]
Diffusion Map
& .2425 & \textbf{.2505} & .2459
& .2303 & .2467 & .2450 & .2567 \\
\addlinespace[3pt]
$\Lng(0.6)$
& .2429 & \textbf{.2505} & .2459
& .2320 & \textbf{.2460} & .2469 & \textbf{.2556} \\
\bottomrule
\end{tabular}
\end{table}

\section{Conclusion}\label{sec:discussion}

We investigate the role of the degree parameter $\alpha$ across three generalized Laplacian variants: shifted, graph-difference, and extra-normalized. Our results show that the effect of $\alpha$ depends critically on how the generalized Laplacian is constructed. In particular, the extra-normalized variant is less sensitive to the choice of $\alpha$ and exhibits more stable performance in our experiments. Across synthetic and real datasets, $\Lng$ achieves ordering errors comparable to other spectral methods and lower than manifold-learning baselines, while attaining the best or near-best phase estimation accuracy.

Despite these benefits, our method still relies on the underlying closed-curve geometry. Abrupt geometric changes or near-overlaps between distant parts of the trajectory may distort the graph's neighborhood structure and compromise phase recovery. Our current theory provides a conditional perturbation-to-phase guarantee, while bounding the perturbation between the observed and expected Laplacians remains necessary for full consistency. Moreover, our theoretical analysis assumes uniform sampling, although the additional experiments in Appendix~\ref{appsec:nonUnif} suggest that $\Lng$ can still achieve accurate recovery under non-uniform sampling with a suitable choice of $\alpha$. Future work will focus on completing the consistency analysis, extending it to non-uniform sampling, and developing data-adaptive choices of $\alpha$.

\section*{Acknowledgements}
Ruizi Wu acknowledges support through the NUS Research Scholarship and conference travel funding from the Department of Statistics and Data Science, Faculty of Science, National University of Singapore. 
Dr. Wanjie Wang was supported by the Singapore Ministry of Education Academic Research Fund Tier 1 under Grant A-8001451-00-00 and Grant A-8003581-00-00. 
We thank the Institute for Mathematical Sciences (IMS) at the National University of Singapore for organizing the workshop Parametric and Nonparametric Network Inference, which initiated this project, and the organizers of the Second Workshop on Mathematical Data Science for facilitating further discussions.

\bibliographystyle{plain}
\bibliography{bib}

\newpage

\appendix

\section*{Appendix}

This section is organized into five appendices. 
Appendix~\ref{appsec:spectralseriation} presents the intuition and illustration of the baseline spectral seriation algorithm. 
Appendix~\ref{appsec:techdetails} provides all the assumptions, technical lemmas, theoretical results, and proofs. 
Implementation details for Section \ref{sec:simulation} and \ref{sec:real} can be found in Appendix~\ref{appsec:simudetails}. 
Appendix~\ref{appsec:addsimu} presents additional simulations under the main setup. 
Appendix~\ref{appsec:nonUnif} investigates non-uniform sampling of latent phases.

\section{Baseline spectral seriation}\label{appsec:spectralseriation}
In this section, we introduce the underlying mathematical intuition of the baseline spectral seriation method. 

Since the latent trajectory is periodic, the collection $\{X(t):0\leq t<2\pi\}$ forms a closed one-dimensional manifold $\M$ embedded in a potentially high-dimensional ambient space. The Laplace-Beltrami operator on $\M$ has a constant eigenfunction with eigenvalue $0$, and its first nontrivial eigenspace is spanned by the two phase functions $\cos(t)$ and $\sin(t)$, both with eigenvalue $1$. More generally, the nontrivial eigenfunctions are $\cos(kt)$ and $\sin(kt)$, with eigenvalues $k^2$ for integers $k\geq 1$. Such spectral structure gives the ideal population representation
\begin{equation}\label{eq:ideal-embedding}
    (\Psi_n)_i := \left(\cos(t_i),\sin(t_i)\right),
    \qquad i\in[n],
\end{equation}
where $(\Psi_n)_i$ denotes the $i$th row of $\Psi_n$. Hence, each sample is represented by a point on the unit circle, and recovering the temporal label $t_i$ is equivalent to recovering its angular coordinate. The graph-based method therefore aims to approximate this two-dimensional phase eigenspace from noisy observations.

\begin{figure}[h]
    \centering
    \includegraphics[width=0.8\linewidth]{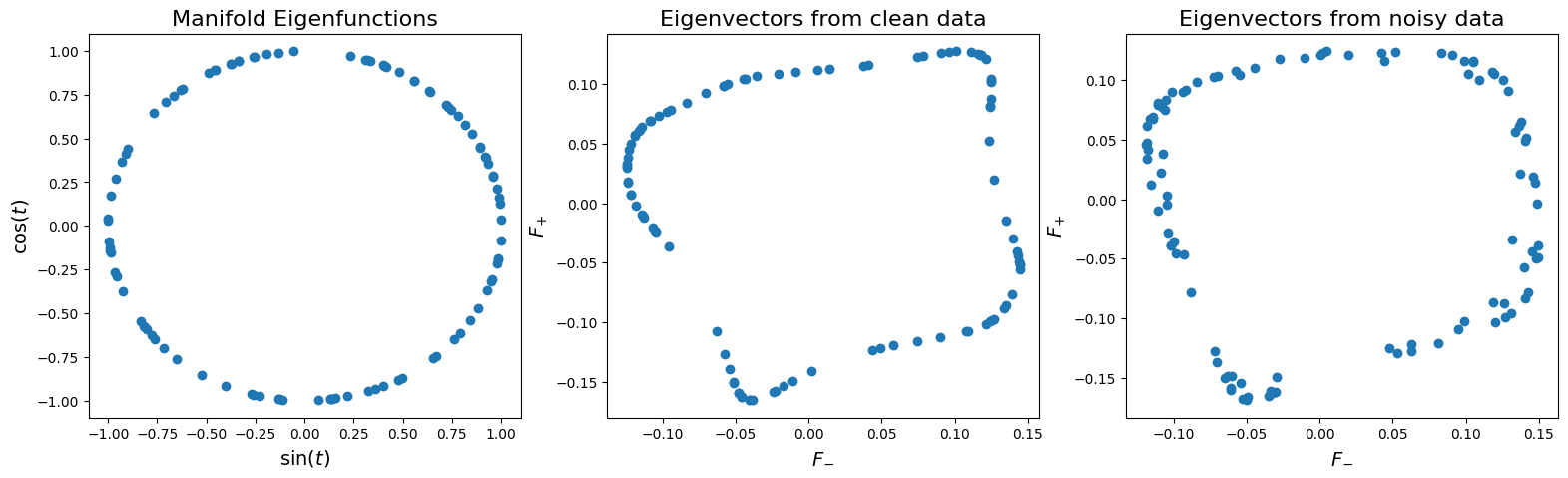}
    \caption{Spectral embedding for periodic manifold data. From left to right, the panels show the population manifold eigenfunctions $\cos(t)$ versus $\sin(t)$, the graph Laplacian eigenvectors $F_+$ versus $F_-$ computed from clean data, and the corresponding eigenvectors computed from noisy data, respectively.}
    \label{figN2026:Intuition}
\end{figure}

As a result, spectral seriation uses $F_+$ and $F_-$ as the data version of $\cos(t)$ and $\sin(t)$. There will be two sources of error: the error from the manifold Laplacian and eigenfunctions to the Laplacian $L$ and eigenvectors based on clean data $X(t_i)$, and the error from the clean data to the noisy data $Z_i$. In Figure~\ref{figN2026:Intuition}, we illustrate how the eigenfunctions contain the information about temporal phases $t_i$, and the effects of two error sources. 

In~Figure \ref{figN2026:Intuition}, the left panel shows the ideal phase representation $(\cos(t),\sin(t))$ on the unit circle. The middle panel demonstrates that in the clean-data case, the graph eigenvector embedding recovers the same circular geometry up to rotation and scaling. The right panel shows that under noise, the embedding is perturbed but still approximately preserves the circular phase structure. Consequently, temporal label recovery can be reduced to angular recovery in the two-dimensional spectral embedding.

\section{Assumptions, technical lemmas and results, and proofs}\label{appsec:techdetails}

Throughout this appendix, we use the notation introduced in Section~\ref{sec:theory} unless otherwise specified. We first state the model assumptions in Section \ref{appsec:assump}. In Section~\ref{appsec:manifoldconvergence}, we establish that the reference-kernel eigenspace converges to the sampled trigonometric eigenspace. Building on this result, we prove Proposition~\ref{Main_prop} in Appendix~\ref{appsec:prop} and Theorem~\ref{Main_theo} in Appendix~\ref{appsec:theo}.

\subsection{Model assumptions}\label{appsec:assump}
To state the theoretical results, we first specify the model assumptions for $X(t)$ and the noise.

\begin{assumption}\label{aspt:setup}
    The observed data points $Z_i = X(t_i) + e_i$ with $i\in[n]$ satisfy the following:
    \begin{enumerate}
        \item[1)] The true data map $X:[0,2\pi] \to \reals^p$ is smooth with $X(0)=X(2\pi)$, and $\|X'(t)\|=1$.
        \item[2)] The manifold $\M$ formed by $\{X(t): 0 \le t \le 2\pi\}$ is isometrically embedded in $\reals^p$.
        \item[3)] The timestamps $t_i$ are independent and uniformly distributed on $[0, 2\pi]$.
        \item[4)] The noise vectors $e_i\in\reals^p$ are independent and distributed as $\mathcal{N}(\zero, \sigma^2 I_p)$, where $\sigma>0$ is unknown.
    \end{enumerate}
\end{assumption}

\begin{assumption}\label{aspt:egap}
    To streamline the theoretical results, we impose the following conditions.
    \begin{enumerate}
        \item[1)] Let $\lambda^K_i\ge0$ and $\lambda^H_i\ge0$ be the $i$th smallest eigenvalue of $L^\star_{\alpha,K}$ and $L^\star_{\alpha,H}$, respectively. There exist constants $\gamma_{1}, \gamma_{2}>0$ such that $\lambda^K_2 \ge \gamma_1\tau$ and $\lambda_4( \tau^{-1} L^\star_{\alpha,K} ) \ge \frac{1}{2}+\gamma_2$. Define $c_{\rm{gap}} := \min(\gamma_1, \frac{\gamma_2}{2})$. Then $\min\left\{ \lambda^H_2 - \lambda^H_1, \lambda^H_4 - \lambda^H_3 \right\} \ge c_{\rm{gap}}\tau$.
        \item[2)] The bandwidth and sample size satisfy that $\tau \to 0$, $\frac{\sqrt{\log{n}}}{n^{1/2}\tau^{5/4}} \to 0$, and $\frac{\rho_{h,\tau,p}}{n\tau^{3/2}} \to 0$.
    \end{enumerate}
\end{assumption}

    Assumption \ref{aspt:egap} above guarantees that the target eigenspace is well separated from the eigenspaces spanned by the other eigenvectors. The assumption of $\frac{\sqrt{\log{n}}}{n^{1/2}\tau^{5/4}} \to 0$ is equivalent to $h \gg (\log(n)/n)^{1/5}$. If $\sigma=O(h)$ and the first two bandwidth assumptions are satisfied, the condition of $\frac{\rho_{h,\tau,p}}{n\tau^{3/2}} \to 0$ follows automatically.

\subsection{Convergence from the reference kernel to manifold}\label{appsec:manifoldconvergence}

To analyze the expected kernel matrix $H$, we introduce a reference Gaussian kernel matrix $K \in \reals^{n\times n}$ with entries
\begin{equation}
    K_{ij}
    = C_{h,\tau}
    \exp\left\{ -\frac{\|X(t_i)-X(t_j)\|_2^2}{2\tau} \right\},
    \qquad i,j\in[n].
\end{equation}
Recall that $C_{h,\tau}$, $\tau$, and $\rho_{h,\tau,p}$ are defined in Section~\ref{sec:theory}. By construction, $K_{ij} = H_{ij}$ for $i\neq j$, while $K_{ii}=C_{h,\tau}$. Hence, $ H=K+C_{h,\tau}\rho_{h,\tau,p}I_n$.

In this subsection, we establish convergence of the eigenspace spanned by $F_{\alpha,K}$ to the sampled trigonometric eigenspace spanned by $\Phi_n$. We begin by defining the associated continuum and empirical operators.

\begin{definition}
\label{def:Singer_functions}
    For any nonnegative continuous kernel function $\kappa:[0,\infty)\to\reals_+$, let
    \begin{equation*}
        \kappa_\tau(x, y) 
        := \kappa\left( \frac{\|x-y\|_{\reals^p}}{\tau^{1/2}} \right),
        \qquad x, \, y\in\M
    \end{equation*}
    Denote $p(y)$ as the sampling density on $\M$. Define the associated continuum quantities as
    \begin{align*}
        p_\tau(x)
        = \int_{\M}\kappa_\tau p(y)\,dV(y),
        \qquad
        \kappa_{\tau,\alpha}(x,y)
        = \frac{\kappa_\tau(x,y)}{p_\tau(x)^\alpha p_\tau(y)^\alpha}, \\
        d_{\tau,\alpha}(x)
        = \int_{\M}\kappa_{\tau,\alpha}(x,y)p(y)\,dV(y),
        \qquad
        m_{\tau,\alpha}(x,y)
        = \frac{\kappa_{\tau,\alpha}(x,y)}{d_{\tau,\alpha}(x)}.
    \end{align*}
    Similarly, define the empirical counterparts as
    \begin{align*}
        \widehat{p}_{\tau,n}(x)
        = \frac{1}{n}\sum_{\ell=1}^n\kappa_\tau(x,x_\ell),
        \qquad
        \widehat{\kappa}_{\tau,\alpha,n}(x,y)
        = \frac{\kappa_\tau(x,y)}{\widehat{p}_{\tau,n}(x)^\alpha \widehat{p}_{\tau,n}(y)^\alpha}, \\
        \widehat{d}_{\tau,\alpha,n}(x)
        = \frac{1}{n}\sum_{j=1}^n \widehat\kappa_{\tau,\alpha,n}(x,x_j),
        \qquad
        \widehat{m}_{\tau,\alpha,n}(x,y)
        = \frac{\widehat\kappa_{\tau,\alpha,n}(x,y)}{\widehat{d}_{\tau,\alpha,n}(x)}.
    \end{align*}
\end{definition}

For any smooth scalar function $f:\mathcal{M}\to\mathbb{R}$,
define the associated continuum and empirical operators as
\begin{align}
    \T_{\tau,\alpha}f(x) 
    & = \int_{\M} m_{\tau,\alpha}(x,y)f(y)p(y)\,dV(y) 
    \label{eq:M_mani_operatopr} \\
    \widehat{\T}_{\tau,\alpha,n}f(x)
    & = \frac{1}{n}\sum_{j=1}^n \widehat{m}_{\tau,\alpha,n}(x,x_j)f(x_j).
    \label{eq:M_pop_operator}
\end{align}

The lemma below provides the pointwise approximation of the reference graph operator by the Laplace--Beltrami operator, thereby linking the discrete construction to the continuum geometry.
\begin{lemma}
\label{lem:pw_converge}
Assume that Assumption \ref{aspt:setup} holds and denote by $x_i = X(t_i)$. Let $f\in C^4(\M)$ and write
    \begin{equation*}
        f_n=(f(x_1),\ldots,f(x_n))^\top.
    \end{equation*}
    For $\alpha\in[0,1]$, further assume that $\tau\to0$ and $\frac{\sqrt{\log(n)}}{n^{1/2} \tau^{5/4}} \to 0$. Then we have that with probability at least $1-n^{-2}$, for all $i\in[n]$,
    \begin{equation*}
        (P_{\alpha, K} f_n-f_n)_i
        = (\T_{\tau,\alpha}f-f)(x_i) + O\left( \frac{\sqrt{\log(n)}}{n^{1/2}\tau^{1/4}} \right),
    \end{equation*}
    where $\T_{\tau,\alpha}$ is defined in \eqref{eq:M_mani_operatopr}. Moreover, it holds that 
    \begin{equation*}
        (\T_{\tau,\alpha}f-f)(x_i)
        = \frac{\tau}{2} \Delta_{\M} f(x_i) + O(\tau^2).
    \end{equation*}
    Consequently, with probability at least $1-n^{-2}$, we have that 
    \begin{equation*}
        \tau^{-1} (P_{\alpha, K}f_n-f_n)_i
        = \frac{1}{2} \Delta_{\M}f(x_i) + O(\tau) + O\left( \frac{\sqrt{\log(n)}}{n^{1/2}\tau^{5/4}} \right)
    \end{equation*}
    uniformly over the sample points.
\end{lemma}

\textit{Proof}. 
We will apply Theorem 5.3 and Corollary 5.1 of \cite{Singer2016}. In their notation, our setting corresponds to the trivial line bundle, i.e., $q=1$ and $g_{ij}=1$. It remains to verify that their empirical normalized operator coincides with our matrix $P_{\alpha, K}$; once this identification is made, the results cited above will be applicable directly.

With $x_i = X(t_i)$, rewrite the reference kernel matrix $K$ as
\begin{equation*}
    K_{ij}
    = C_{h, \tau} \exp\left\{-\frac{\|x_i -x_j\|^2}{2\tau}\right\}.
\end{equation*}
With such notation, the empirical quantities in Definition \ref{def:Singer_functions} can be rewritten as
\begin{align*}
    & \widehat{p}_{\tau,n}(x_i)
    = \frac{D_K(i)}{n}, 
    \qquad 
    \widehat{K}_{\tau,\alpha,n}(x_i,x_j)
    = n^{2\alpha}D_K(i)^{-\alpha}K_{ij}D_K(j)^{-\alpha}
    = n^{2\alpha}\widetilde{K}_{\alpha, ij}, \\
    & \widehat{d}_{\tau,\alpha,n}(x_i)
    = n^{2\alpha-1}\widetilde{D}_{\alpha, K}(i),
    \qquad 
    \frac{1}{n} \widehat{M}_{\tau,\alpha,n}(x_i,x_j)
    = \frac{\widetilde{K}_{\alpha, ij}}{\widetilde{D}_{\alpha, K}(i)}
    = P_{\alpha, K}(i,j).
\end{align*}
Consequently, for each smooth $f$, it holds that 
\begin{equation*}
    (P_{\alpha, K}f_n-f_n)_i
    = \widehat{\T}_{\tau,\alpha,n}f(x_i)-f(x_i).
\end{equation*}
By resorting to Theorem 5.3 of \cite{Singer2016}, with probability at least $1-n^{-2}$, we have that for all $i\in[n]$,
\begin{equation*}
    (P_{\alpha, K}f_n-f_n)_i
    = (\T_{\tau,\alpha}f-f)(x_i) + O\left( \frac{\sqrt{\log(n)}}{n^{1/2}\tau^{1/4}} \right)
\end{equation*}
for each fixed $\alpha \in [0, 1]$. In our setting, the case of $\alpha=0$ satisfies the same conclusion as the case of $0<\alpha\le 1$, since $\M$ has no boundary contribution. Moreover, under Assumption \ref{aspt:setup}, the induced sampling density on $\M$ is constant. Hence, the density-drift term in Theorem 5.2 of \cite{Singer2016} vanishes, which gives that 
\begin{equation*}
    (\T_{\tau,\alpha}f-f)(x_i)
    = \frac{\tau}{2} \Delta_{\M}f(x_i)+O(\tau^2), 
    \qquad 0\le \alpha \le 1.
\end{equation*}
Therefore, combining these estimates and dividing by $\tau$ yield that 
\begin{equation*}
    \tau^{-1} (P_{\alpha, K}f_n-f_n)_i
    = \frac{1}{2} \Delta_{\M}f(x_i) + O(\tau) + O\left( \frac{\sqrt{\log(n)}}{n^{1/2}\tau^{5/4}} \right).
\end{equation*}
This completes the proof of Lemma \ref{lem:pw_converge}.

The lemma below verifies that the additional $\alpha$-normalization does not introduce non-negligible degree heterogeneity, which will be needed to compare weighted and unweighted eigenspaces.
\begin{lemma}
\label{lem:degree_K}
Assume that all the conditions of Lemma \ref{lem:pw_converge} are satisfied. Then we have that with probability at least $1 - n^{-2}$, for any fixed $\alpha\in[0,1]$,
    \begin{equation*}
        \max_{i\in[n]} \left| \frac{\widetilde{D}_{\alpha, K}(i)}{\bar{\widetilde{D}}_{\alpha, K}} - 1 \right|
        = O \left( (1 + \alpha)\left( \tau + \sqrt{\frac{\log(n)}{n\tau^{1/2}}} \right) \right)
        = o(1),
    \end{equation*}
    where $\bar{\widetilde{D}}_{\alpha, K} = \frac{1}{n}\sum_{i=1}^n \widetilde{D}_{\alpha, K}(i)$.
\end{lemma}

\textit{Proof}. 
We first show that the unnormalized reference degrees $D_K(i)$ are uniformly flat. Recall that
\begin{equation*}
    D_K(i) = \sum_{j=1}^nK_{ij},
    \qquad
    K_{ij}
    = C_{h,\tau} \exp \left\{ -\frac{\|X(t_i)-X(t_j)\|^2}{2\tau} \right\}.
\end{equation*}
Conditional on $t_i$, random variables $\{K_{ij}:j\ne i\}$ are independent and identically distributed (i.i.d.). For fixed $t_i=t$, it holds that 
\begin{equation*}
    \E[K_{ij}| t_i=t]
    = C_{h,\tau}\frac1{2\pi} \int_0^{2\pi} \exp\left\{ -\frac{\|X(t)-X(s)\|^2}{2\tau} \right\}\,ds.
\end{equation*}
Since $X$ is a smooth closed embedded curve with $\|X'(t)\|=1$, there exists some $r_0>0$ such that for all $t\in[0, 2\pi]$ and 
$|u|\le r_0$,
\begin{equation*}
    \|X(t+u)-X(t)\|^2 = u^2+O(u^4).
\end{equation*}
This expansion follows from Taylor's theorem. For $r_0<|u|\le \pi$, we have that 
$$\|X(t+u) - X(t)\|\geq \delta >0.$$ 
It follows that the contribution from $|u|>r_0$ is $O(e^{-c/\tau})$. Hence, after a change of variable $u=s-t$ and using the periodicity, we can deduce that 
\begin{align*}
    \int_0^{2\pi} \exp\left\{ -\frac{\|X(t) - X(s)\|^2}{2\tau} \right\}\,ds
    & = \int_{-\pi}^\pi \exp\left\{ -\frac{\|X(t+u) - X(t)\|^2}{2\tau} \right\}\,du \\
    & = \int_{|u|\le r_0} \exp \left\{-\frac{u^2 + O(u^4)}{2\tau}\right\}\,du + O(e^{-c/\tau}).
\end{align*}

With a change of variable $u = \sqrt{\tau}z$, the Gaussian moment and tail estimates give that 
\begin{equation*}
    \int_0^{2\pi} \exp\left\{ -\frac{\|X(t) - X(s)\|^2}{2\tau} \right\}\,ds 
    = \sqrt{2\pi\tau} (1 + O(\tau)).
\end{equation*}
Taking the supremum over $t\in[0, 2\pi]$ on both sides, it holds that 
\begin{equation*}
    \E[K_{ij}| t_i]
    = \frac{1}{\sqrt{2\pi}} C_{h,\tau} \tau^{1/2}(1+O(\tau)).
\end{equation*}
Moreover, we have that $0\le K_{ij}\le C_{h,\tau}$. Conditional on $t_i=t$, we have that for any $j\neq i$,
\begin{equation*}
    \E[K_{ij}^2| t_i=t]
    = C_{h,\tau}^2\frac1{2\pi} \int_0^{2\pi} \exp\left\{ -\frac{\|X(t)-X(s)\|^2}{\tau} \right\}\,ds.
\end{equation*}
By the same local Gaussian-integral estimate, we can show that 
\begin{equation*}
    \sup_{t\in[0,2\pi]} \int_0^{2\pi} \exp\left\{ -\frac{\|X(t)-X(s)\|^2}{\tau} \right\}\,ds
    = \sqrt{\pi\tau} (1 + O(\tau)),
\end{equation*}
Consequently, we can obtain that 
\begin{equation*}
    \E[K_{ij}^2| t_i]
    \le C C_{h,\tau}^2\tau^{1/2}.
\end{equation*}

Observe that for fixed $t_i$ and $j\neq i$, random variables $K_{ij} - \E[K_{ij}|t]$ are independent, have mean zero, and satisfy that $|K_{ij} - \E[K_{ij}|t]| \le CC_{h,\tau}$. Moreover, it holds that 
\begin{equation*}
    \sum_{j\neq i} \E[(K_{ij} - \E[K_{ij}|t])^2 | t_i] \le \sum_{j\neq i} \E[K_{ij}^2 | t_i] \leq CnC_{h,\tau}^2 \tau^{1/2}.
\end{equation*}
An application of Bernstein's inequality leads to 
\begin{equation*}
    \left| \sum_{j\ne i}K_{ij} - (n-1) \E[K_{ij}| t_i] \right|
    \le C C_{h,\tau} \left( \sqrt{n\tau^{1/2}\log(n)} + \log(n) \right)
\end{equation*}
with probability at least $1-n^{-3}$. Since $\tau\to0$ and $\frac{\sqrt{\log(n)}}{n^{1/2} \tau^{5/4}} \to 0$, term $\log(n)(n\tau^{1/2})^{-1} \to 0$ automatically. Thus, the second term is negligible relative to $nC_{h,\tau}\tau^{1/2}$. Then a union bound over $i\in[n]$ results in 
\begin{equation*}
    \max_{i\in[n]}
    \left| \sum_{j\ne i}K_{ij} - (n-1)\E[K_{ij}| t_i] \right| 
    \le C C_{h,\tau} \sqrt{n\tau^{1/2}\log(n)}
\end{equation*}
with probability at least $1-n^{-2}$. Since $K_{ii}=C_{h,\tau}$, combining the preceding bounds yields that 
\begin{equation*}
    D_K(i)
    = nC_{h,\tau} \frac{\tau^{1/2}}{\sqrt{2\pi}} ( 1 + O\left( \delta_n \right) ),
    \qquad
    \delta_n = \tau + \sqrt{\frac{\log(n)}{n\tau^{1/2}}}
\end{equation*}
with probability at least $1-n^{-2}$. By the assumptions on $\tau$ and $n$, it follows that 
\begin{equation*}
    D_K(i)^{-\alpha}
    = \left( nC_{h,\tau} \frac{\tau^{1/2}}{\sqrt{2\pi}} \right)^{-\alpha} ( 1 + O \left( \alpha \delta_n \right) ).
\end{equation*}

Further, we can show that 
\begin{align*}
    \widetilde{D}_{\alpha, K}(i)
    & = \sum_{j=1}^n D_K(i)^{-\alpha} K_{ij} D_K(j)^{-\alpha}  \\
    & = \left( nC_{h,\tau} \frac{\tau^{1/2}}{\sqrt{2\pi}} \right)^{-2\alpha} ( 1 + O \left( \alpha \delta_n \right) ) \sum_{j=1}^nK_{ij}.
\end{align*}
Using again the uniform expansion for $D_K(i)=\sum_jK_{ij}$, it holds that 
\begin{equation*}
    \widetilde{D}_{\alpha, K}(i)
    = \left( nC_{h,\tau} \frac{\tau^{1/2}}{\sqrt{2\pi}} \right)^{1-2\alpha} ( 1 + O \left( (1 + \alpha) \delta_n \right) ), \qquad \forall i\in[n].
\end{equation*}
Averaging the last expression over $i$ leads to 
\begin{equation*}
    \bar{\widetilde{D}}_{\alpha, K}
    = \left( nC_{h,\tau} \frac{\tau^{1/2}}{\sqrt{2\pi}} \right)^{1-2\alpha} ( 1 + O \left( (1 + \alpha) \delta_n \right) ).
\end{equation*}
Therefore, we can obtain that 
\begin{equation*}
    \max_{i\in[n]} \left| \frac{\widetilde{D}_{\alpha, K}(i)}{\bar{\widetilde{D}}_{\alpha, K}} - 1 \right|
    = O \left( (1 + \alpha) \delta_n \right),
\end{equation*}
where the right-hand side is $o(1)$ under the stated bandwidth condition. This concludes the proof of Lemma \ref{lem:degree_K}.

\begin{definition}
\label{def:eigenspace}
    The matrix $\Psi_n$ collects the two continuum eigenfunctions
    $\cos(t)$ and $\sin(t)$ evaluated at the sample points
    \begin{equation}\label{eq:Psi_n}
        \Psi_n=\left[ 
        \begin{array}{cc}
            \cos(t_1) & \sin(t_1) \\
            \cos(t_2) & \sin(t_2) \\ 
            \vdots & \vdots \\
            \cos(t_n) & \sin(t_n) \\
        \end{array} \right].
    \end{equation}
    Denote the reference basis corresponding to the empirical trigonometric space $\rm{span}\{(\cos t_i)_i,(\sin t_i)_i\}$ as 
    \begin{equation}\label{eq:ThetaSpace}
        \Theta_{\alpha, n}
        := \widetilde{D}_{\alpha, K}^{1/2} \Psi_n \left( \Psi_n^\top \widetilde{D}_{\alpha, K} \Psi_n \right)^{-1/2}
    \end{equation}
    with $(\Theta_{\alpha, n})^\top \Theta_{\alpha, n} = I_2$. Let $q_{\alpha, n} :=  \frac{\widetilde{D}_{\alpha, K}^{1/2} \one} {\| \widetilde{D}_{\alpha, K}^{1/2} \one \|_2}$. The orthonormalized projection of $\Theta_{\alpha, n}$ onto $q_{\alpha, n}^\perp$ is given by 
    \begin{equation}\label{eq:Theta_orthospace}
        \Theta_{\alpha, n}^\perp
        =  (I - q_{\alpha, n} (q_{\alpha, n})^\top) \Theta_{\alpha, n} \left( (\Theta_{\alpha, n})^\top(I-q_{\alpha, n}(q_{\alpha, n})^\top)\Theta_{\alpha, n} \right)^{-1/2}.
    \end{equation}
    Define
    \begin{equation*}
        \Phi_n 
        := \Psi_n(\Psi_n^\top \Psi_n)^{-1/2}.
    \end{equation*}
    Consequently, $\Phi_n$ has orthonormal columns and spans the empirical trigonometric space
    \begin{equation*}
        \rm{span}\{(\cos t_i)_{i=1}^n,(\sin t_i)_{i=1}^n\}.
    \end{equation*}
\end{definition}

We next translate the pointwise operator approximation in Lemma \ref{lem:pw_converge} into an approximate eigen-relation for the empirical trigonometric basis. This step provides a bridge from pointwise convergence to eigenspace convergence.

\begin{lemma}
\label{lem:approxi_eigen}
Assume that all the conditions of Lemma \ref{lem:pw_converge} are satisfied. Then we have that with probability $1 - O(n^{-2})$,
    \begin{equation*}
        \left\| \left( \tau^{-1} L_{\alpha, K}^\star - \frac{1}{2} I \right)\Theta_{\alpha, n} \right\|_F
        = O\left( \tau + \frac{\sqrt{\log(n)}}{n^{1/2}\tau^{5/4}} \right),
    \end{equation*}
    where $\Theta_{\alpha, n}$ is defined in \eqref{eq:ThetaSpace}.
\end{lemma}

\textit{Proof}. 
Denote by 
\begin{equation*}
    r_n := \tau + \frac{\sqrt{\log(n)}}{n^{1/2}\tau^{5/4}}.
\end{equation*}
By invoking Lemma \ref{lem:pw_converge}, it holds that for each fixed $f\in C^4(\M)$,
\begin{equation*}
    \tau^{-1}(P_{\alpha, K} f_n-f_n)_i
    = \frac{1}{2} \Delta_M f(x_i)+O(r_n)
\end{equation*}
with probability at least $1 - n^{-2}$. We will apply this relation to $f(X(t))=\cos(t)$. Since $X$ is unit speed with $X(0)=X(2\pi)$, it follows that 
\begin{equation*}
    \Delta_{\M} f(X(t))
    = \frac{d^2}{dt^2}\cos(t)
    = -\cos(t)
    = -f(X(t)).
\end{equation*}
Hence, we have that for all $i\in[n]$,
\begin{equation*}
    \tau^{-1}(P_{\alpha, K}f_n-f_n)_i
    = -\frac{1}{2}f(X(t_i)) + O(r_n)
    = -\frac{1}{2}\cos(t_i) + O(r_n),
\end{equation*}
where $f_n:=(f(X(t_1)), \ldots, f(X(t_n))^\top = (\cos(t_1), \ldots, \cos(t_n))^\top$. Equivalently, we have that 
\begin{equation*}
    \left(\tau^{-1}L_{\alpha, K} f_n-\frac{1}{2} f_n\right)_i
    = O(r_n).
\end{equation*}
An application of the same argument with $f(t)=\sin(t)$ leads to 
\begin{equation*}
    \left( \tau^{-1}L_{\alpha, K} g_n-\frac{1}{2} g_n \right)_i
    = O(r_n),
    \qquad
    g_n
    = (\sin(t_1), \ldots, \sin(t_n))^\top.
\end{equation*}

Let $E_{\alpha, n} := \left( \tau^{-1} L_{\alpha, K} - \frac{1}{2} I \right)\Psi_n$. The $i$th row of $E_{\alpha, n}$ can be written as
\begin{equation*}
    (E_{\alpha, n})_i 
    = \left( \left( \tau^{-1}L_{\alpha, K} f_n - \frac{1}{2}f_n \right)_i, \left( \tau^{-1}L_{\alpha, K} g_n - \frac{1}{2}g_n \right)_i \right).
\end{equation*}
Then it holds that $\max_{i\in[n]} \|(E_{\alpha, n})_i\|_2 = O(r_n)$ with probability at least $1 - 2n^{-2}$. Since
\begin{equation*}
    L_{\alpha, K}^\star
    = \widetilde{D}_{\alpha, K}^{1/2}L_{\alpha, K}\widetilde{D}_{\alpha, K}^{-1/2},
\end{equation*}
we have that 
\begin{equation*}
    \tau^{-1} L_{\alpha, K}^\star
    = \widetilde{D}_{\alpha, K}^{1/2} \left(\tau^{-1} L_{\alpha, K}\right) \widetilde{D}_{\alpha, K}^{-1/2}.
\end{equation*}
Thus, we can deduce that 
\begin{align*}
    \left( \tau^{-1} L_{\alpha, K}^\star - \frac{1}{2} I \right) \Theta_{\alpha, n}
    & = \widetilde{D}_{\alpha, K}^{1/2} \left( \tau^{-1} L_{\alpha, K} - \frac{1}{2} I \right) \Psi_n \left( \Psi_n^\top \widetilde{D}_{\alpha, K} \Psi_n \right)^{-1/2} \\
    & = \widetilde{D}_{\alpha, K}^{1/2} E_{\alpha, n} \left( \Psi_n^\top \widetilde{D}_{\alpha, K} \Psi_n \right)^{-1/2}.
\end{align*}

We now bound the two factors above separately. Since $\max_{i\in[n]} \|(E_{\alpha, n})_i \|_2 \le C r_n$ with probability at least $1 - 2n^{-2}$, it holds that 
\begin{equation*}
    \|\widetilde{D}_{\alpha, K}^{1/2} E_{\alpha, n} \|_F
    = \sqrt{ \sum_{i=1}^n
    \widetilde{D}_{\alpha, K}(i) \| (E_{\alpha, n})_i \|_2^2}
    \le \sqrt{ C r_n^2 \sum_{i=1}^n \widetilde{D}_{\alpha, K}(i) }
    = C r_n \sqrt{ n \bar{\widetilde{D}}_{\alpha, K} }
\end{equation*}
with probability at least $1-n^{-2}$. An application of Lemma \ref{lem:degree_K} shows that with probability at least $1 - n^{-2}$,
\begin{equation*}
    \Psi_n^\top \widetilde{D}_{\alpha, K} \Psi_n
    = \bar{\widetilde{D}}_{\alpha, K} \Psi_n^\top \Psi_n + o(n\bar{\widetilde{D}}_{\alpha, K}).
\end{equation*}
Denote by $a_n = \frac{1}{n} \sum_{i=1}^n \cos(2t_i)$ and $b_n = \frac{1}{n} \sum_{i=1}^n \sin(2t_i)$. Then it follows that 
\begin{equation*}
    \frac{1}{n} \Psi_n^\top \Psi_n
    = \frac{1}{2} I_2
    + \frac{1}{2}
    \begin{pmatrix}
        a_n & b_n\\
        b_n & -a_n
    \end{pmatrix}.
\end{equation*}
Notice that since $t_i\stackrel{i.i.d.}{\sim}\rm{Unif}[0,2\pi]$, random variables $\cos(2t_i)$ and $\sin(2t_i)$ are independent, mean-zero, and bounded in $[-1,1]$. Then an application of Hoeffding's inequality gives that 
\begin{equation*}
    \mathbb{P} \left( |a_n| > \frac{1}{2\sqrt2} \right)
    \le 2e^{-n/16},
    \qquad
    \mathbb{P} \left( |b_n| > \frac{1}{2\sqrt2} \right)
    \le 2e^{-n/16}.
\end{equation*}
Consequently, we have that with probability at least $1 - 4e^{-n/16}$, $\sqrt{a_n^2+b_n^2}\le \frac{1}{2}$ and thus 
\begin{equation*}
    \lambda_{\min} \left( \frac{1}{n} \Psi_n^\top \Psi_n \right)
    = \frac{1}{2} - \frac{1}{2} \sqrt{a_n^2 + b_n^2}
    \ge \frac{1}{4}.
\end{equation*}
Equivalently, we have that $\lambda_{\min}(\Psi_n^\top \Psi_n) \ge \frac{n}{4}$. 

Further, by resorting to Weyl's inequality, we can deduce that 
\begin{align*}
    \lambda_{\min}(\Psi_n^\top\widetilde{D}_{\alpha, K}\Psi_n)
    & \ge \bar{\widetilde{D}}_{\alpha, K} \lambda_{\min}(\Psi_n^\top\Psi_n) - \left\| \Psi_n^\top \widetilde{D}_{\alpha, K} \Psi_n - \bar{\widetilde{D}}_{\alpha, K} \Psi_n^\top \Psi_n \right\|_2  \\
    & \ge \frac{1}{4} n\bar{\widetilde{D}}_{\alpha, K} - o(n\bar{\widetilde{D}}_{\alpha, K}) \\
    & \ge cn\bar{\widetilde{D}}_{\alpha, K}
\end{align*}
with probability at least $1-4e^{-n/16}-n^{-2}$. Therefore, combining the two bounds yields that with probability $1 - O(n^{-2})$,
\begin{equation*}
    \left\| \left( \tau^{-1} L_{\alpha, K}^\star - \frac{1}{2} I \right)\Theta_{\alpha, n} \right\|_F
    \le \|\widetilde{D}_{\alpha, K}^{1/2} E_{\alpha, n} \|_F \left\| \left( \Psi_n^\top \widetilde{D}_{\alpha, K} \Psi_n \right)^{-1/2} \right\|_2 
    \le C r_n.
\end{equation*}
Since $r_n = \tau + \frac{\sqrt{\log(n)}}{n^{1/2} \tau^{5/4}} = o(1)$ under the bandwidth condition of Lemma \ref{lem:degree_K}, the desired conclusion follows. This completes the proof of Lemma \ref{lem:approxi_eigen}.

Using the approximate eigen-relation from Lemma \ref{lem:approxi_eigen}, we next obtain an upper bound for the first nontrivial eigenvalues of the rescaled reference graph Laplacian.

\begin{lemma}\label{lem:Bound_K_evals}
Assume that all the conditions of Lemma \ref{lem:pw_converge} are satisfied. Let $L_{\alpha, n} := \tau^{-1} L_{\alpha, K}^\star$, and assume further that $0=\lambda_1(L_{\alpha, n})\le \lambda_2(L_{\alpha, n})\le \lambda_3(L_{\alpha, n})\le\cdots$. Then we have that with probability $1-O(n^{-2})$,
    \begin{equation*}
        \lambda_2(L_{\alpha, n}), \, \lambda_3(L_{\alpha, n})\le \frac{1}{2} + O \left( (1 + \alpha) \left( \tau + \frac{\sqrt{\log(n)}}{n^{1/2}\tau^{5/4}} \right) \right).
    \end{equation*}
    Equivalently, we have that with probability $1-O(n^{-2})$,
    \begin{equation*}
        \lambda_2(L_{\alpha, K}^\star), \, \lambda_3(L_{\alpha, K}^\star)\le \frac{\tau}{2} + O \left( (1 + \alpha) \left( \tau + \frac{\sqrt{\log(n)}}{n^{1/2}\tau^{5/4}} \right) \right).
    \end{equation*}
\end{lemma}

\textit{Proof}. 
Recall that $\Theta_{\alpha, n}$, $q_{\alpha, n}$, and $\Theta_{\alpha, n}^\perp$ defined in \eqref{eq:ThetaSpace} and \eqref{eq:Theta_orthospace}. Also, recall that
\begin{equation*}
    P_{\alpha, K}^\star
    :=
    \widetilde{D}_{\alpha, K}^{-1/2} \widetilde{K}_{\alpha} \widetilde{D}_{\alpha, K}^{-1/2},
    \qquad
    L_{\alpha, K}^\star=I-P_{\alpha, K}^\star,
\end{equation*}
It holds that $P_{\alpha, K}^\star q_{\alpha, n}=q_{\alpha, n}$ and thus $L_{\alpha, n}q_{\alpha, n}=0$. Moreover, since the Gaussian kernel is strictly positive, $\widetilde{K}_{\alpha}$ has strictly positive entries. By the Perron--Frobenius theorem, the eigenvalue $1$ of $P_{\alpha, K}^\star$ is simple. Hence, the zero eigenspace of $L_{\alpha, n}$ is $\rm{span}\{q_{\alpha, n}\}$. In view of Lemma \ref{lem:approxi_eigen}, we have that with probability $1 - O(n^{-2})$,
\begin{equation*}
    \left\| \left( L_{\alpha, n} - \frac{1}{2} I \right)\Theta_{\alpha, n} \right\|_F
    = O\left( \tau + \frac{\sqrt{\log(n)}}{n^{1/2}\tau^{5/4}} \right).
\end{equation*}
It follows from Lemma \ref{lem:degree_K} that with probability $1 - n^{-2}$,
\begin{equation*}
    \max_{i\in[n]}
    \left|
    \frac{\widetilde{D}_{\alpha, K}(i)}{\bar{\widetilde{D}}_{\alpha, K}}-1
    \right|
    = O ((1+\alpha)\delta_n).
\end{equation*}

Applying Hoeffding's inequality to the entries of $\Psi_n$ and $\Psi_n^\top\Psi_n$, followed by a union bound, gives that with probability $1-O(n^{-2})$,
\begin{equation*}
    \frac{1}{n}\one^\top \Psi_n
    = O \left(\sqrt{\frac{\log(n)}{n}}\right),
    \qquad
    \frac{1}{n}\Psi_n^\top\Psi_n
    = \frac{1}{2} I_2 + O\left(\sqrt{\frac{\log(n)}{n}}\right).
\end{equation*}
This entails that 
\begin{align*}
    \frac{1}{n\bar{\widetilde{D}}_{\alpha, K}}
    \one^\top \widetilde{D}_{\alpha, K}\Psi_n
    = O\left((1+\alpha)\delta_n + \sqrt{\frac{\log(n)}{n}}\right), \\
    \frac{1} {n\bar{\widetilde{D}}_{\alpha, K}} \Psi_n^\top \widetilde{D}_{\alpha, K}\Psi_n
    = \frac{1}{2} I_2 + O\left((1+\alpha)\delta_n + \sqrt{\frac{\log(n)}{n}}\right).
\end{align*}
Consequently, we have that with probability $1-O(n^{-2})$,
\begin{equation*}
    (q_{\alpha, n})^\top \Theta_{\alpha, n}
    = \frac{\one^\top \widetilde{D}_{\alpha, K} \Psi_n}{\left( \one^\top \widetilde{D}_{\alpha, K} \one \right)^{1/2}} \left( \Psi_n^\top \widetilde{D}_{\alpha, K} \Psi_n \right)^{-1/2} 
    = O\left( (1+\alpha) \delta_n \right).
\end{equation*}

Let us recall the definition
\begin{equation*}
    \Theta_{\alpha, n}^\perp
    =
    (I-q_{\alpha, n}(q_{\alpha, n})^\top)\Theta_{\alpha, n}
    \left[
    (\Theta_{\alpha, n})^\top(I-q_{\alpha, n}(q_{\alpha, n})^\top)\Theta_{\alpha, n}
    \right]^{-1/2}.
\end{equation*}
It holds that 
\begin{equation*}
    (\Theta_{\alpha, n})^\top (I-q_{\alpha, n}(q_{\alpha, n})^\top) \Theta_{\alpha, n}
    = I_2 + O\left( (1+\alpha) \delta_n \right).
\end{equation*}
Under our bandwidth conditions, we have that $O\left( (1+\alpha) \delta_n \right)=o(1)$. Hence, with probability $1-O(n^{-2})$, $\Theta_{\alpha, n}^\perp$ is well-defined, has orthonormal columns, satisfies that $\rm{span}(\Theta_{\alpha, n}^\perp)\subset q_{\alpha, n}^\perp$, and has $\rm{dim}(\rm{span}(\Theta_{\alpha, n}^\perp))=2$. It remains to verify that the residual estimate is preserved after projection. It follows from $L_{\alpha, n}q_{\alpha, n}=0$ that 
\begin{equation*}
    \left( L_{\alpha, n} - \frac{1}{2} I \right) q_{\alpha, n}
    = -\frac{1}{2} q_{\alpha, n}.
\end{equation*}
Consequently, we can deduce that 
\begin{align*}
    \left( L_{\alpha, n} - \frac{1}{2} I \right) (I - q_{\alpha, n} (q_{\alpha, n})^\top) \Theta_{\alpha, n}
    & = \left( L_{\alpha, n} - \frac{1}{2} I \right) \Theta_{\alpha, n} - \left( L_{\alpha, n} - \frac{1}{2} I \right) q_{\alpha, n} (q_{\alpha, n})^\top \Theta_{\alpha, n}  \\
    & = \left( L_{\alpha, n} - \frac{1}{2} I \right) \Theta_{\alpha, n}
    + \frac{1}{2} q_{\alpha, n} (q_{\alpha, n})^\top \Theta_{\alpha, n}.
\end{align*}

Taking the Frobenius norms and invoking Lemma \ref{lem:approxi_eigen}, we can show that with probability $1-O(n^{-2})$,
\begin{equation*}
    \left\| \left( L_{\alpha, n} - \frac{1}{2} I \right)(I - q_{\alpha, n} (q_{\alpha, n})^\top) \Theta_{\alpha, n} \right\|_F
    = O\left( \tau + \frac{\sqrt{\log n}}{n^{1/2}\tau^{5/4}} + (1+\alpha) \delta_n \right).
\end{equation*}
In light of the condition $\tau \to 0$, it holds that 
\begin{equation*}
    \left\| \left( L_{\alpha, n} - \frac{1}{2} I \right) \Theta_{\alpha, n}^\perp \right\|_F
    = O\left( \tau + \frac{\sqrt{\log n}}{n^{1/2}\tau^{5/4}} + (1+\alpha) \delta_n \right)
    = O \left( (1 + \alpha) \left( \tau + \frac{\sqrt{\log(n)}}{n^{1/2}\tau^{5/4}} \right) \right).
\end{equation*}
For any $u\in \rm{span}(\Theta_{\alpha, n}^\perp)$ with $\|u\|_2=1$, we can write $u=\Theta_{\alpha, n}^\perp a$ with $\|a\|_2=1$. Then for all $u\in \rm{span}(\Theta_{\alpha, n}^\perp)$ with $\|u\|_2=1$, it follows that 
\begin{align*}
    \left| u^\top \left( L_{\alpha, n} - \frac{1}{2} I \right) u \right|
    & \le \left\| \left( L_{\alpha, n} - \frac{1}{2} I \right) \Theta_{\alpha, n}^\perp \right\|_2 \\
    & \le \left\| \left( L_{\alpha, n} - \frac{1}{2} I \right) \Theta_{\alpha, n}^\perp \right\|_F \\
    & = O \left( (1 + \alpha) \left( \tau + \frac{\sqrt{\log(n)}}{n^{1/2}\tau^{5/4}} \right) \right).
\end{align*}
Thus, we can obtain that 
\begin{equation*}
    \max_{\substack{u\in \rm{span}(\Theta_{\alpha, n}^\perp)\\ \|u\|_2=1}} u^\top L_{\alpha, n} u
    \le \frac{1}{2} + O \left( (1 + \alpha) \left( \tau + \frac{\sqrt{\log(n)}}{n^{1/2}\tau^{5/4}} \right) \right).
\end{equation*}

Finally, by the min-max theorem on $q_{\alpha, n}^\perp$, it holds that 
\begin{equation*}
    \lambda_3(L_{\alpha, n})
    = \min_{\substack{S\subset q_{\alpha, n}^\perp\\ \dim(S)=2}}
    \;
    \max_{\substack{u\in S\\ \|u\|_2=1}}
    u^\top L_{\alpha, n} u.
\end{equation*}
Using $S=\rm{span}(\Theta_{\alpha, n}^\perp)$ gives that
\begin{equation*}
    \lambda_3(L_{\alpha, n})
    \le \frac{1}{2} + O \left( (1 + \alpha) \left( \tau + \frac{\sqrt{\log(n)}}{n^{1/2}\tau^{5/4}} \right) \right).
\end{equation*}
Since $\lambda_2(L_{\alpha, n}) \le \lambda_3(L_{\alpha, n})$, we also have that 
\begin{equation*}
    \lambda_2(L_{\alpha, n})
    \le \frac{1}{2} + O \left( (1 + \alpha) \left( \tau + \frac{\sqrt{\log(n)}}{n^{1/2}\tau^{5/4}} \right) \right).
\end{equation*}
Therefore, the final statement follows from $L_{\alpha, n}=\tau^{-1}L_{\alpha, K}^\star$. This concludes the proof of Lemma \ref{lem:Bound_K_evals}.

We are now ready to establish the main result of this subsection: the convergence of the reference eigenspace to the empirical trigonometric space spanned by $(\cos(t_i))_{i=1}^n$ and $(\sin(t_i))_{i=1}^n$.

\begin{proposition}[Conditional reference eigenspace convergence]
\label{prop:SpanKConverge}
Assume that Assumptions \ref{aspt:setup}--\ref{aspt:egap} hold. Then we have that with probability $1-O(n^{-2})$,
    \begin{equation*}
        \|F_{\alpha, K} F_{\alpha, K}^\top - \Theta_{\alpha,n}^{\perp}(\Theta_{\alpha,n}^{\perp})^\top \|_2
        = O \left( (1 + \alpha) \left( \tau + \frac{\sqrt{\log(n)}}{n^{1/2}\tau^{5/4}} \right) \right).
    \end{equation*}
    Consequently, we have that with probability $1-O(n^{-2})$,
    \begin{equation*}
        \|F_{\alpha, K} F_{\alpha, K}^\top - \Phi_n \Phi_n^\top\|_2
        = O \left( (1 + \alpha) \left( \tau + \frac{\sqrt{\log(n)}}{n^{1/2}\tau^{5/4}} \right) \right).
    \end{equation*}
    Equivalently, there exists some $Q\in \mathbb{O}(2)$ such that with probability $1-O(n^{-2})$,
    \begin{equation*}
        \|F_{\alpha, K}Q - \Phi_n\|_F
        = O \left( (1 + \alpha) \left( \tau + \frac{\sqrt{\log(n)}}{n^{1/2}\tau^{5/4}} \right) \right).
    \end{equation*}
\end{proposition}

\textit{Proof}. 
Denote by $L_{\alpha, n}=\tau^{-1}L_{\alpha, K}^\star$. Then $L_{\alpha, n}$ is symmetric and has the same eigenspaces as $L_{\alpha,K}^\star$. We denote its eigenvalues as
\begin{equation*}
    0=\lambda_1(L_{\alpha, n})\le \lambda_2(L_{\alpha, n})\le \lambda_3(L_{\alpha, n})
    \le \lambda_4(L_{\alpha, n})\le \cdots \le \lambda_n(L_{\alpha, n}).
\end{equation*}
In view of the proof of Lemma \ref{lem:Bound_K_evals}, it holds that 
\begin{equation*}
    r_n 
    := \left\| \left( L_{\alpha, n} - \frac{1}{2} I \right) \Theta_{\alpha,n}^{\perp} \right\|_F
    = O \left( (1 + \alpha) \left( \tau + \frac{\sqrt{\log(n)}}{n^{1/2}\tau^{5/4}} \right) \right)
\end{equation*}
with probability $1-O(n^{-2})$. Let $F_\perp$ be an orthonormal eigenbasis for the direct sum of eigenspaces of $L_{\alpha, n}$ corresponding to $\lambda_4(L_{\alpha, n}),\ldots,\lambda_n(L_{\alpha, n})$. Since $\Theta_{\alpha,n}^{\perp}\subset q_{\alpha,n}^{\perp}$, we can write
\begin{equation*}
    \Theta_{\alpha,n}^{\perp}
    = F_{\alpha, K} C_n + F_\perp S_n
\end{equation*}
for some matrices $C_n\in\mathbb R^{2\times 2}$ and $S_n\in\mathbb R^{(n-3)\times 2}$. Since $\Theta_{\alpha,n}^{\perp}$ has orthonormal columns, we have that 
\begin{equation*}
    C_n^\top C_n+S_n^\top S_n=I_2.
\end{equation*}
By Assumption \ref{aspt:egap}, there exists a fixed constant $\gamma>0$ such that $\lambda_4(L_{\alpha, n}) \geq \frac{1}{2} + \gamma$. Hence, we can deduce that 
\begin{align*}
    r_n^2
    & = \left\| \left( L_{\alpha, n} - \frac{1}{2} I \right) \Theta_{\alpha,n}^{\perp} \right\|_F^2 \\
    & = \left\| \left( L_{\alpha, n} - \frac{1}{2} I\right) F_{\alpha, K} C_n \right\|_F^2 + \left\| \left( L_{\alpha, n} - \frac{1}{2} I \right) F_\perp S_n \right\|_F^2 \\
    & \geq \gamma^2\|S_n\|_F^2.
\end{align*}
It follows that 
\begin{equation*}
    \|S_n\|_F
    \le \frac{r_n}{\gamma}
    = O \left( (1 + \alpha) \left( \tau + \frac{\sqrt{\log(n)}}{n^{1/2}\tau^{5/4}} \right) \right).
\end{equation*}
Since
\begin{equation*}
    (I - F_{\alpha, K} F_{\alpha, K}^\top) \Theta_{\alpha,n}^{\perp} 
    = F_\perp S_n,
\end{equation*}
we can obtain that 
\begin{equation*}
    \|(I - F_{\alpha, K} F_{\alpha, K}^\top) \Theta_{\alpha,n}^{\perp}\|_2
    \le \|S_n\|_F
    = O \left( (1 + \alpha) \left( \tau + \frac{\sqrt{\log(n)}}{n^{1/2}\tau^{5/4}} \right) \right).
\end{equation*}
Since both $F_{\alpha, K}$ and $\Theta_{\alpha,n}^{\perp}$ have two orthonormal columns, this entails that 
\begin{equation*}
    \left\| F_{\alpha, K}F_{\alpha, K}^\top - \Theta_{\alpha,n}^{\perp} (\Theta_{\alpha,n}^{\perp})^\top \right\|_2
    = O \left( (1 + \alpha) \left( \tau + \frac{\sqrt{\log(n)}}{n^{1/2}\tau^{5/4}} \right) \right).
\end{equation*}

It remains to relate $\rm{span}(\Theta_{\alpha,n}^{\perp})$ to $\rm{span}(\Phi_n)$. By Lemma \ref{lem:degree_K}, we have that with probability $1-O(n^{-2})$,
\begin{equation*}
    \bar{\widetilde{D}}_{\alpha,K}^{-1/2}
    \widetilde{D}_{\alpha,K}^{1/2}
    = I + O ((1 + \alpha)\delta_n).
\end{equation*}
Since right multiplication by an invertible $2\times2$ matrix does not change the column space, it holds that 
\begin{equation*}
    \rm{span}(\Theta_{\alpha,n})
    = \rm{span}(\widetilde{D}_{\alpha,K}^{1/2} \Psi_n).
\end{equation*}
Moreover, in view of $\Psi_n=\Phi_n(\Psi_n^\top\Psi_n)^{1/2}$, we have that 
\begin{equation*}
    \rm{span}(\widetilde{D}_{\alpha,K}^{1/2}\Psi_n)
    = \rm{span}(\widetilde{D}_{\alpha,K}^{1/2}\Phi_n).
\end{equation*}
It follows that 
\begin{equation*}
    \rm{span}(\Theta_{\alpha,n})
    = \rm{span}\!\left(
    \left( \bar{\widetilde{D}}_{\alpha,K}^{-1/2} \widetilde{D}_{\alpha,K}^{1/2} \right)\Phi_n \right).
\end{equation*}
Thus, $\rm{span}(\Theta_{\alpha,n})$ is an $O((1+\alpha)\delta_n)$-perturbation of $\rm{span}(\Phi_n)$. Then it holds that 
\begin{equation*}
    \| \Theta_{\alpha,n} \Theta_{\alpha,n}^{\top} - \Phi_n \Phi_n^{\top} \|_2
    = O((1+\alpha)\delta_n).
\end{equation*}
Moreover, in the proof of Lemma \ref{lem:Bound_K_evals}, we have that $(q_{\alpha,n})^{\top}\Theta_{\alpha,n}=O\left( (1+\alpha) \delta_n \right)$. Consequently, removing the $q_{\alpha,n}$-component changes the associated projector by at most $O\left( (1+\alpha) \delta_n \right)$
\begin{equation*}
    \left\| \Theta_{\alpha,n}^{\perp} (\Theta_{\alpha,n}^{\perp})^\top - \Theta_{\alpha,n}\Theta_{\alpha,n}^\top \right\|_2
    = O\left( (1+\alpha) \delta_n \right).
\end{equation*}

Combining the two expressions above gives that 
\begin{equation*}
    \left\| \Theta_{\alpha,n}^{\perp} (\Theta_{\alpha,n}^{\perp})^\top - \Phi_n\Phi_n^\top \right\|_2
    = O\left( (1+\alpha) \delta_n \right).
\end{equation*}
An application of the triangle inequality shows that 
\begin{align*}
    \|F_{\alpha, K} F_{\alpha, K}^\top - \Phi_n \Phi_n^\top\|_2
    & \le \left\| F_{\alpha, K} F_{\alpha, K}^\top - \Theta_{\alpha,n}^{\perp} (\Theta_{\alpha,n}^{\perp})^\top \right\|_2 + \left\| \Theta_{\alpha,n}^{\perp} (\Theta_{\alpha,n}^{\perp})^\top - \Phi_n\Phi_n^\top \right\|_2 \\
    & = O \left( (1 + \alpha) \left( \tau + \frac{\sqrt{\log(n)}}{n^{1/2}\tau^{5/4}} \right) \right).
\end{align*}
Finally, since $F_{\alpha, K}$ and $\Phi_n$ are both $n\times 2$ matrices with orthonormal columns, the convergence of the rank-two projectors implies the convergence of orthonormal bases up to a $2\times2$ orthogonal transformation. Therefore, there exists some $Q\in \mathbb{O}(2)$ such that
\begin{equation*}
    \|F_{\alpha, K}Q - \Phi_n\|_F 
    = O \left( (1 + \alpha) \left( \tau + \frac{\sqrt{\log(n)}}{n^{1/2}\tau^{5/4}} \right) \right).
\end{equation*}
This completes the proof of Proposition \ref{prop:SpanKConverge}.

\subsection{From reference kernel to clean Laplacian}

In the previous subsection, we have established the spectral convergence for the reference kernel Laplacian based on $K$. We now transfer this convergence to the $H$-based Laplacian by first controlling the perturbation induced by the diagonal difference between $H$ and $K$.

\begin{lemma}
\label{lem:diagH}
    Let $\alpha\in[0,1]$ and $H=K+R$, where $R$ is diagonal. Define
    \begin{equation*}
        \epsilon_r
        := \max_{i\in[n]} \frac{|R_{ii}|}{D_K(i)} + \max_{i\in[n]} \frac{D_K(i)^{-2\alpha}|R_{ii}|}{\widetilde{D}_{\alpha,K}(i)}.
    \end{equation*}
    Assume that $(1 + \alpha)\epsilon_r\le c_\epsilon$ for a sufficiently small absolute constant $c_\epsilon>0$. Then we have that with probability $1-O(n^{-2})$,
    \begin{equation*}
        \|L^\star_{\alpha,H} - L^\star_{\alpha,K}\|_2
        \le C(1 + \alpha) \epsilon_r \left( 1 + (1 + \alpha) \left( \tau + \sqrt{\frac{\log(n)}{n\tau^{1/2}}} \right) \right).
    \end{equation*}
\end{lemma}

\textit{Proof}. 
Let $\xi_i = \frac{R_{ii}}{D_K(i)}$. It holds that 
\begin{equation*}
    D_H(i)
    = D_K(i)+R_{ii}
    = D_K(i)(1+\xi_i).
\end{equation*}
From the definition of $\epsilon_r$, we see that $\max_{i\in[n]}|\xi_i|\le \epsilon_r$. Since $(1 + \alpha)\epsilon_r\le 1/2$, we have that $|\xi_i|\le 1/2$. By the mean-value theorem, it follows that 
\begin{equation*}
    \left|(1 + \xi_i)^{-\alpha} - 1\right|
    \le C (1 + \alpha)|\xi_i|
    \le C (1 + \alpha) \epsilon_r.
\end{equation*}
Hence, we can show that 
\begin{equation*}
    D_H(i)^{-\alpha}
    = D_K(i)^{-\alpha}(1 + O((1 + \alpha)\epsilon_r)).
\end{equation*}
For $i\neq j$, since $H_{ij}=K_{ij}$, we have that 
\begin{equation*}
    \widetilde{H}_{\alpha,ij}
    = D_H(i)^{-\alpha} K_{ij} D_H(j)^{-\alpha}
    = \widetilde{K}_{\alpha,ij}(1 + O((1 + \alpha) \epsilon_r)).
\end{equation*}
Thus, we can obtain that 
\begin{equation*}
    |\widetilde{H}_{\alpha,ij} - \widetilde{K}_{\alpha,ij}|
    \le C (1 + \alpha) \epsilon_r \widetilde{K}_{\alpha,ij}.
\end{equation*}

For the diagonal entries, it holds that 
\begin{equation*}
    \widetilde{H}_{\alpha, ii} - \widetilde{K}_{\alpha, ii}
    = (D_H(i)^{-2\alpha} - D_K(i)^{-2\alpha})H_{ii} + D_K(i)^{-2\alpha} (H_{ii} - K_{ii}).
\end{equation*}
It follows from
\begin{equation*}
    D_H(i)^{-2\alpha} = D_K(i)^{-2\alpha} (1 + O((1 + \alpha) \epsilon_r))
\end{equation*}
that 
\begin{equation*}
    |(D_H(i)^{-2\alpha} - D_K(i)^{-2\alpha})H_{ii}| 
    \le C (1 + \alpha) \epsilon_r D_K(i)^{-2\alpha} H_{ii}
    \le C (1 + \alpha) \epsilon_r (\widetilde{K}_{\alpha, ii} + D_K(i)^{-2\alpha}|R_{ii}|).
\end{equation*}
Hence, we have that 
\begin{equation*}
    |\widetilde{H}_{\alpha,ii} - \widetilde{K}_{\alpha,ii}|
    \le C (1 + \alpha) \epsilon_r \widetilde{K}_{\alpha,ii} + (1 + C (1 + \alpha) \epsilon_r) D_K(i)^{-2\alpha}|R_{ii}|.
\end{equation*}
By the definition of $\epsilon_r$, it holds that $D_K(i)^{-2\alpha}|R_{ii}| \le \epsilon_r \widetilde{D}_{\alpha, K}(i)$. After summing over $j$, we can deduce that 
\begin{equation*}
    \sum_{j=1}^n |\widetilde{H}_{\alpha,ij} - \widetilde{K}_{\alpha,ij}|
    \le C (1 + \alpha) \epsilon_r \widetilde{D}_{\alpha,K}(i).
\end{equation*}
Consequently, we have that 
\begin{equation*}
    |\widetilde{D}_{\alpha,H}(i) - \widetilde{D}_{\alpha,K}(i)|
    \le C (1 + \alpha) \epsilon_r \widetilde{D}_{\alpha,K}(i)
\end{equation*}
and thus $\widetilde{D}_{\alpha,H}(i) = \widetilde{D}_{\alpha,K}(i)(1 + O(\alpha \epsilon_r))$. 

Decomposing $|P^\star_{\alpha,H}(i,j)-P^\star_{\alpha,K}(i,j)|$ gives that 
\begin{align*}
    |P^\star_{\alpha,H}(i,j) - P^\star_{\alpha,K}(i,j)|
    & \le \frac{|\widetilde{H}_{\alpha,ij} - \widetilde{K}_{\alpha,ij}|}{\sqrt{\widetilde{D}_{\alpha,H}(i) \widetilde{D}_{\alpha,H}(j)}} \\
    & \quad + \widetilde{K}_{\alpha,ij} \left| \frac{1}{\sqrt{\widetilde{D}_{\alpha,H}(i) \widetilde{D}_{\alpha,H}(j)}} - \frac{1}{\sqrt{\widetilde{D}_{\alpha,K}(i) \widetilde{D}_{\alpha,K}(j)}} \right|.
\end{align*}
For sufficiently small $c_\epsilon$, it holds that $C(1 + \alpha)\epsilon_r\le 1/2$. Then we have that 
\begin{align*}
    |P^\star_{\alpha,H}(i,j) - P^\star_{\alpha,K}(i,j)|
    & \le (1 + C(1 + \alpha) \epsilon_r) \frac{|\widetilde{H}_{\alpha,ij} - \widetilde{K}_{\alpha,ij}|}{\sqrt{\widetilde{D}_{\alpha,K}(i) \widetilde{D}_{\alpha,K}(j)}} \\
    & \qquad + C (1 + \alpha) \epsilon_r \frac{\widetilde{K}_{\alpha,ij}}{\sqrt{\widetilde{D}_{\alpha,K}(i) \widetilde{D}_{\alpha,K}(j)}}.
\end{align*}
It follows from Lemma \ref{lem:degree_K} that with probability $1-O(n^{-2})$,
\begin{equation*}
    \max_{i\in[n]} \left| \frac{\widetilde{D}_{\alpha,K}(i)}{\bar{\widetilde{D}}_{\alpha,K}} - 1\right|
    \le C(1+\alpha)\delta_n,
\end{equation*}
where $\delta_n = \tau + \sqrt{\frac{\log n}{n\tau^{1/2}}}$. 

For $n$ sufficiently large, it holds that $C(1+\alpha)\delta_n<1/2$. For each $i$, we have that 
\begin{equation*}
    (1 - C(1+\alpha)\delta_n) \bar{\widetilde{D}}_{\alpha,K}
    \le \widetilde{D}_{\alpha,K}(i)
    \le (1 + C(1+\alpha)\delta_n) \bar{\widetilde{D}}_{\alpha,K}.
\end{equation*}
For any $i,j$, it follows that 
\begin{equation*}
    \widetilde{D}_{\alpha,K}(j)
    \ge (1 - C(1+\alpha)\delta_n) \bar{\widetilde{D}}_{\alpha,K}
    \ge \frac{1 - C(1+\alpha)\delta_n}{1 + C(1+\alpha)\delta_n} \widetilde{D}_{\alpha,K}(i).
\end{equation*}
Consequently, we can show that 
\begin{equation*}
    \frac{1}{\sqrt{\widetilde{D}_{\alpha,K}(i) \widetilde{D}_{\alpha,K}(j)}}
    \le (1 + C(1+\alpha) \delta_n) \frac{1}{\widetilde{D}_{\alpha,K}(i)}.
\end{equation*}
Summing over $j$ yields that 
\begin{equation*}
    \max_{i\in[n]} \sum_j |P^\star_{\alpha,H}(i,j) - P^\star_{\alpha,K}(i,j)|
    \le C(1 + \alpha)\epsilon_r (1 + (1+\alpha)\delta_n).
\end{equation*}

Since \(P^\star_{\alpha,H}-P^\star_{\alpha,K}\) is symmetric, Schur's test gives that 
\begin{equation*}
    \|P^\star_{\alpha,H}-P^\star_{\alpha,K}\|_2
    \le C(1 + \alpha)\epsilon_r (1 + (1+\alpha)\delta_n).
\end{equation*}
Consequently, we can obtain that 
\begin{equation*}
    \|L^\star_{\alpha,H} - L^\star_{\alpha,K}\|_2
    \le C(1 + \alpha) \epsilon_r (1 + (1 + \alpha) \delta_n).
\end{equation*}
This concludes the proof of Lemma \ref{lem:diagH}.

It remains to verify that the diagonal perturbation size appearing in Lemma \ref{lem:diagH} is negligible under the stated bandwidth conditions.

\begin{lemma}
\label{lem:exactH}
Assume that all the conditions of Lemma \ref{lem:pw_converge} are satisfied. Then we have that with probability $1 - O(n^{-2})$,
    \begin{equation*}
        \epsilon_r
        := \max_{i\in[n]} \frac{|R_{ii}|}{D_K(i)} + \max_{i\in[n]} \frac{D_K(i)^{-2\alpha}|R_{ii}|}{\widetilde{D}_{\alpha,K}(i)}
        \le C \frac{\rho_{h,\tau,p}}{n\tau^{1/2}}.
    \end{equation*}
    Consequently, if $\frac{\rho_{h,\tau,p}}{n\tau^{3/2}}\to 0$, we have that  $\epsilon_r=o(\tau)$.
\end{lemma}

\textit{Proof}. 
By Lemma \ref{lem:degree_K} and its proof, we have that with probability $1 - O(n^{-2})$, for all $i\in[n]$,
\begin{equation*}
    D_K(i)
    = \frac{nC_{h,\tau}\tau^{1/2}}{\sqrt{2\pi}} (1+O(\delta_n))
\end{equation*}
and
\begin{equation*}
    \widetilde{D}_{\alpha,K}(i)
    = \left( \frac{nC_{h,\tau}\tau^{1/2}}{\sqrt{2\pi}} \right)^{1-2\alpha} (1+O((1+\alpha)\delta_n)).
\end{equation*}
Since $\tau \to 0$ and $\frac{\sqrt{\log(n)}}{n^{1/2} \tau^{5/4}} \to 0$, for $n$ sufficiently large it holds that $(1+\alpha)\delta_n \le c_\delta$ for a sufficiently small absolute constant $c_\delta>0$. These estimates imply the uniform lower bounds
\begin{equation*}
    D_K(i) 
    \ge c nC_{h,\tau} \tau^{1/2}
\end{equation*}
and
\begin{equation*}
    \widetilde{D}_{\alpha,K}(i)
    \ge c \left( nC_{h,\tau}\tau^{1/2} \right)^{1-2\alpha}.
\end{equation*}
It follows from $H - K = R = C_{h,\tau} \rho_{h,\tau,p}I_n$ that $|R_{ii}| = C_{h,\tau} \rho_{h,\tau,p}$ for each $i$. Hence, we have that 
\begin{equation*}
    \max_{i\in[n]} \frac{|R_{ii}|}{D_K(i)}
    \le C \frac{C_{h,\tau}\rho_{h,\tau,p}}{nC_{h,\tau}\tau^{1/2}}
    = C\frac{\rho_{h,\tau,p}}{n\tau^{1/2}}.
\end{equation*}

It remains to bound the second term in $\epsilon_r$. Using the same lower bounds, we can deduce that 
\begin{equation*}
    \max_{i\in[n]} \frac{D_K(i)^{-2\alpha}|R_{ii}|}{\widetilde{D}_{\alpha,K}(i)}
    \le C \frac{(nC_{h,\tau}\tau^{1/2})^{-2\alpha} C_{h,\tau}\rho_{h,\tau,p}}{(nC_{h,\tau}\tau^{1/2})^{1-2\alpha}}
    = C\frac{\rho_{h,\tau,p}}{n\tau^{1/2}}.
\end{equation*}
Combining the two bounds yields that 
\begin{equation*}
    \epsilon_r
    = \max_{i\in[n]}\frac{|R_{ii}|}{D_K(i)} + \max_{i\in[n]} \frac{D_K(i)^{-2\alpha}|R_{ii}|}{\widetilde{D}_{\alpha,K}(i)}
    \le C\frac{\rho_{h,\tau,p}}{n\tau^{1/2}}.
\end{equation*}
Finally, we can obtain that 
\begin{equation*}
    \frac{\epsilon_r}{\tau}
    \le C\frac{\rho_{h,\tau,p}}{n\tau^{3/2}},
\end{equation*}
Therefore, the condition of $\frac{\rho_{h,\tau,p}}{n\tau^{3/2}}\to0$ entails that $\epsilon_r=o(\tau)$, as required. This completes the proof of Lemma \ref{lem:exactH}.

\subsection{Proof of Proposition \ref{Main_prop}}\label{appsec:prop}

It follows from the triangle inequality that 
\begin{equation*}
    \|F_{\alpha,H} F_{\alpha,H}^{\top} -\Phi_n \Phi_n^{\top}\|_2
    \le \|F_{\alpha,H} F_{\alpha,H}^{\top} - F_{\alpha,K} F_{\alpha,K}^{\top}\|_2 + \|F_{\alpha,K} F_{\alpha,K}^{\top} - \Phi_n \Phi_n^{\top}\|_2.
\end{equation*}
We first control the deviation of the reference kernel. By resorting to Proposition \ref{prop:SpanKConverge}, it holds that with probability $1 - O(n^{-2})$,
\begin{equation*}
    \|F_{\alpha,K} F_{\alpha,K}^{\top} - \Phi_n \Phi_n^{\top}\|_2
    \le C \left( (1 + \alpha) \left( \tau + \frac{\sqrt{\log(n)}}{n^{1/2}\tau^{5/4}} \right) \right).
\end{equation*}
It remains to transfer the eigenspace from $K$ to $H$. By Lemma \ref{lem:exactH}, we have that with probability $1 - O(n^{-2})$, $\epsilon_r \le C\frac{\rho_{h,\tau,p}}{n\tau^{1/2}}$. Since $\frac{\rho_{h,\tau,p}}{n\tau^{3/2}}\to0$, we have that $\epsilon_r = o(\tau)$. Then for all sufficiently large $n$, it holds that $(1+\alpha) \epsilon_r \le c_\epsilon$. An application of Lemma \ref{lem:diagH} leads to 
\begin{equation*}
    \|L^\star_{\alpha,H}-L^\star_{\alpha,K}\|_2
    \le C (1 + \alpha) \epsilon_r \left(1 + (1+\alpha) \delta_n\right).
\end{equation*}
Combining this with the previous bound gives that 
\begin{equation*}
    \|L^\star_{\alpha,H} - L^\star_{\alpha,K}\|_2
    \le C \frac{\rho_{h,\tau,p}}{n\tau^{1/2}} (1+\alpha) \left(1 + (1+\alpha) \delta_n \right).
\end{equation*}
In view of Assumption \ref{aspt:egap}, it holds that $\lambda^K_2 \ge \gamma_1\tau$ and $\lambda_4( \tau^{-1} L^\star_{\alpha,K} ) \ge \frac{1}{2}+\gamma_2$. By Lemma \ref{lem:Bound_K_evals}, for sufficiently large $n$ we have that $\lambda^K_3 \le \left( \frac{1}{2} + \frac{\gamma_2}{2} \right) \tau$ with probability $1-O(n^{-2})$. Let us set $c_{\rm{gap}} = \min(\gamma_1, \gamma_2/2)$. The second--third eigenspace of $L^\star_{\alpha,K}$ is thus separated from the rest of the spectrum by at least $c_{\rm{gap}}\tau$. Since
\begin{equation*}
    \frac{\|L^\star_{\alpha,H} - L^\star_{\alpha,K}\|_2}{c_{\rm{gap}} \tau}
    \le \frac{C}{c_{\rm{gap}}} \frac{\rho_{h,\tau,p}}{n\tau^{3/2}} (1 + \alpha) \left(1 + (1 + \alpha) \delta_n \right) 
    \to 0,
\end{equation*}
the perturbation is smaller than $c_{\rm{gap}}\tau/4$ for all sufficiently large $n$.
Hence, the second--third eigenspace of $L^\star_{\alpha,H}$ is well-defined. 

An application of the Davis--Kahan theorem results in 
\begin{equation*}
    \|F_{\alpha,H} F_{\alpha,H}^{\top} - F_{\alpha,K} F_{\alpha,K}^{\top}\|_2
    \le \frac{C}{c_{\rm{gap}}\tau} \|L^\star_{\alpha,H} - L^\star_{\alpha,K}\|_2.
\end{equation*}
Then it follows that 
\begin{equation*}
    \|F_{\alpha,H} F_{\alpha,H}^{\top} - F_{\alpha,K} F_{\alpha,K}^{\top}\|_2
    \le \frac{C}{c_{\rm{gap}}} \frac{\rho_{h,\tau,p}}{n\tau^{3/2}} (1 + \alpha) \left( 1 + (1 + \alpha) \delta_n \right).
\end{equation*}
Combining the reference term and the perturbation term yields that 
\begin{align*}
    \|F_{\alpha,H} F_{\alpha,H}^{\top} -\Phi_n \Phi_n^{\top}\|_2
    & \le C\left( \tau + \frac{\sqrt{\log n}}{\sqrt n\,\tau^{5/4}} + (1 + \alpha) \delta_n + \frac{1}{c_{\rm{gap}}} \frac{\rho_{h,\tau,p}}{n\tau^{3/2}} (1 + \alpha) \left(1 + (1 + \alpha) \delta_n \right) \right) \\
    & \le C(1 + \alpha) \left( \tau + \frac{\sqrt{\log n}}{\sqrt n\,\tau^{5/4}} + \frac{1}{c_{\rm{gap}}} \frac{\rho_{h,\tau,p}}{n\tau^{3/2}} \right).
\end{align*}
This establishes the projector bound. Finally, since both $F_{\alpha,H}$ and $\Phi_n$ have orthonormal columns, the standard Procrustes equivalence for rank-two projectors gives a $Q\in \mathbb{O}(2)$ such that
\begin{equation*}
    \|F_{\alpha,H}Q - \Phi_n\|_F
    \le C\|F_{\alpha,H} F_{\alpha,H}^{\top} -\Phi_n \Phi_n^{\top}\|_2.
\end{equation*}
Thus, the claimed Frobenius bound follows. This concludes the proof of Proposition \ref{Main_prop}.

\subsection{Proof of Theorem \ref{Main_theo}}\label{appsec:theo}

By the Davis--Kahan theorem and the assumed eigengap, it holds that 
\begin{equation*}
    \|F_{\alpha,S} F_{\alpha,S}^\top - F_{\alpha,H} F_{\alpha,H}^\top\|_2
    \le C \frac{\|L^\star_{\alpha,S} - L^\star_{\alpha,H}\|_2}{c_{\rm{gap}} \tau}
    \le C \frac{U_S}{c_{\rm{gap}}\tau}.
\end{equation*}
Since the eigenspaces are two-dimensional, the standard Procrustes alignment bound for orthonormal bases gives that 
\begin{equation*}
    \inf_{Q\in \mathbb{O}(2)} \|F_{\alpha,S}Q - F_{\alpha,H}\|_F
    \le C\frac{U_S}{c_{\rm{gap}}\tau}.
\end{equation*}
In light of Proposition \ref{Main_prop}, there exists some $Q_1\in \mathbb{O}(2)$ such that with probability $1-O(n^{-2})$,
\begin{equation*}
    \|F_{\alpha,H}Q_1 - \Phi_n\|_F
    \le C(1 + \alpha) \left( \tau + \frac{\sqrt{\log n}}{\sqrt n\,\tau^{5/4}} + \frac{1}{c_{\rm{gap}}} \frac{\rho_{h,\tau,p}}{n\tau^{3/2}} \right).
\end{equation*}
Let $Q_2\in\mathbb{O}(2)$ be a minimizer, or an arbitrarily close minimizer, in the preceding Procrustes bound. Using the fact that the product of two elements in $\mathbb{O}(2)$ remains in $\mathbb{O}(2)$, we can deduce that 
\begin{align*}
    \inf_{Q\in \mathbb{O}(2)} \|F_{\alpha,S}Q - \Phi_n\|_F
    & \le \|(F_{\alpha,S}Q_2 - F_{\alpha,H})Q_1\|_F + \|F_{\alpha,H}Q_1 - \Phi_n\|_F \\
    & \le C \left[ (1 + \alpha) \left( \tau + \frac{\sqrt{\log(n)}}{\sqrt{n}\,\tau^{5/4}} + \frac{1}{c_{\rm{gap}}} \frac{\rho_{h,\tau,p}}{n\tau^{3/2}} \right) + \frac{U_S}{c_{\rm{gap}}\tau} \right].
\end{align*}

Let us define the ideally scaled trigonometric matrix $\Phi_n^0:=\sqrt{\frac{2}{n}}\Psi_n$, and let $G_n:=\frac{2}{n}\Psi_n^\top\Psi_n$. Then we have that 
\begin{equation*}
    \Phi_n
    = \Psi_n( \Psi_n^\top \Psi_n )^{-1/2}
    = \sqrt{\frac{2}{n}} \Psi_n G_n^{-1/2}
    = \Phi_n^0 G_n^{-1/2}.
\end{equation*}
Moreover, it holds that 
\begin{equation*}
    G_n - I_2 =
    \begin{pmatrix}
        a_n & b_n\\
        b_n & -a_n
    \end{pmatrix},
\end{equation*}
where $a_n:=\frac{1}{n}\sum_{i=1}^n\cos(2t_i)$ and $b_n:=\frac{1}{n}\sum_{i=1}^n\sin(2t_i)$. An application of Hoeffding's inequality shows that with probability at least $1-Cn^{-2}$,
\begin{equation*}
    |a_n| + |b_n|
    \le C\sqrt{\frac{\log(n)}{n}}.
\end{equation*}
It follows that 
\begin{equation*}
    \|G_n-I_2\|_2
    \le C\sqrt{\frac{\log(n)}{n}}.
\end{equation*}
Hence, there exists some $n_0$ such that for all $n > n_0$, $\|G_n-I_2\|_2\le 1/2$. Then all eigenvalues of $G_n$ lie in $[1/2,3/2]$. For $f(x)=x^{-1/2}$, we have that $\sup_{x\in[1/2,3/2]} |f'(x)| \le C$. By the spectral decomposition of $G_n$, we can deduce that 
\begin{equation*}
    \|G_n^{-1/2} - I_2\|_2 
    = \max_{j\in\{1, 2\}} |\lambda_j(G_n)^{-1/2} - 1|
    \le C \max_{j\in\{1, 2\}}|\lambda_j(G_n) - 1|
    = C\|G_n - I_2\|_2.
\end{equation*}
Thus, we can obtain that 
\begin{equation*}
    \|G_n^{-1/2} - I_2\|_2
    \le C\sqrt{\frac{\log(n)}{n}}.
\end{equation*}

Since $\|\Phi_n^0\|_F=\sqrt{2}$, it follows that
\begin{equation*}
    \|\Phi_n - \Phi_n^0\|_F
    = \|\Phi_n^0(G_n^{-1/2} - I_2)\|_F
    \le \sqrt{2}\|G_n^{-1/2} - I_2\|_2
    \le C\sqrt{\frac{\log(n)}{n}}.
\end{equation*}
Hence, we can deduce that 
\begin{align*}
    \inf_{Q\in \mathbb{O}(2)} \|F_{\alpha,S}Q - \Phi_n^0\|_F
    & \le \inf_{Q\in \mathbb{O}(2)} \|F_{\alpha,S}Q - \Phi_n \|_F + \|\Phi_n - \Phi_n^0\|_F \\
    & \le C \left[ (1 + \alpha) \left( \tau + \frac{\sqrt{\log(n)}}{\sqrt{n} \tau^{5/4}} + \frac{1}{c_{\rm{gap}}} \frac{\rho_{h,\tau,p}}{n\tau^{3/2}} \right) +
    \frac{U_S}{c_{\rm{gap}}\tau} + \sqrt{\frac{\log(n)}{n}} \right] \\
    & \le C \left[ (1 + \alpha) \left( \tau + \frac{\sqrt{\log(n)}}{\sqrt{n} \tau^{5/4}} + \frac{1}{c_{\rm{gap}}} \frac{\rho_{h,\tau,p}}{n\tau^{3/2}} \right) + \frac{U_S}{c_{\rm{gap}}\tau} \right].
\end{align*}
Denote by 
\begin{equation*}
    R_n 
    = (1 + \alpha) \left( \tau + \frac{\sqrt{\log(n)}}{\sqrt{n} \tau^{5/4}} + \frac{1}{c_{\rm{gap}}} \frac{\rho_{h,\tau,p}}{n\tau^{3/2}} \right) +\frac{U_S}{c_{\rm{gap}}\tau}.
\end{equation*}
We choose $Q\in \mathbb{O}(2)$ such that
\begin{equation*}
\|F_{\alpha,S}Q-\Phi_n^0\|_F\le C R_n.
\end{equation*}

Next, let $\widetilde{E}_n = E_nQ^\top = F_{\alpha,S} - \Phi_n^0Q^\top$. Then it holds that $\|\widetilde{E}_n\|_F = \|E_n\|_F\le C R_n$. Since $Q\in \mathbb{O}(2)$, there exist some $r\in\{-1,1\}$ and $\theta\in[0,2\pi)$ such that the $i$th row of $\Phi_n^0Q^\top$ is
\begin{equation*}
    (\Phi_n^0Q^\top)_i 
    = \sqrt{\frac{2}{n}} \left(\cos(rt_i+\theta),\sin(rt_i+\theta)\right).
\end{equation*}
It follows that 
\begin{equation*}
    (F_{\alpha,S})_{i}
    = \sqrt{\frac{2}{n}} \left(\cos(rt_i+\theta), \sin(rt_i+\theta)\right) + (\widetilde{E}_n)_i.
\end{equation*}
Denote by $r_n = \sqrt{\frac{2}{n}}$ and $u_i = (\cos(rt_i+\theta), \sin(rt_i+\theta) )$. Let $e=(e_1, e_2)\in\mathbb R^2$ be arbitrary. If $\|e\|_2\le r_n/2$, we have that 
\begin{align*}
    \left\| \frac{r_nu_i + e}{\|r_nu_i + e\|_2} - u_i \right\|_2
    & \le \left\| \frac{r_nu_i + e}{\|r_nu_i + e\|_2} - \frac{r_nu_i + e}{r_n} \right\|_2 + \left\| \frac{r_nu_i + e}{r_n} -u_i \right\|_2 \\
    & \le \frac{|r_n - \|r_nu_i + e\|_2|}{r_n} + \frac{\|e\|_2}{r_n} \\
    & \le C \frac{\|e\|_2}{r_n}.
\end{align*}

Denote by $\rm{d} (a, b) = |[a-b]_{2\pi}|$. Observe that the definition of $\rm{Error}(\hat{t}, t)$ is equivalent to
\begin{equation*}
     \min_{r\in\{-1, 1\}, \theta\in[0, 2\pi]} \left( \frac{1}{n} \sum_{i=1}^n \rm{d}(\hat{t}_i, rt_i+\theta)^2 \right)^{\frac{1}{2}}.
\end{equation*}
Since the arc length on the unit circle is bounded by a constant multiple of the Euclidean chord length, it follows that
\begin{equation*}
    \rm{d} \left( rt_i + \theta, \arctan\left( \frac{r_n \sin(rt_i + \theta) + e_2}{r_n \cos(rt_i + \theta) + e_1} \right) \right)
    \le C \frac{\|e\|_2}{r_n}.
\end{equation*}
Squaring both sides and using $r_n^2=2/n$ lead to 
\begin{equation*}
    \rm{d} \left( rt_i+\theta, \arctan\left( \frac{r_n \sin(rt_i + \theta) + e_2}{r_n \cos(rt_i + \theta) + e_1} \right) \right)^2
    \le Cn\|e\|_2^2.
\end{equation*}
If $\|e\|_2>r_n/2$, the trivial bound $\rm{d} (a, b) \le \pi$ for all $a, b\in \reals$ gives that 
\begin{equation*}
    \rm{d} \left( rt_i+\theta, \arctan\left( \frac{r_n \sin(rt_i + \theta) + e_2}{r_n \cos(rt_i + \theta) + e_1} \right) \right)^2
    \le \pi^2
    \le Cn\|e\|_2^2.
\end{equation*}
Consequently, we have that for any $e\in\reals^2$,
\begin{equation*}
    \rm{d} \left( rt_i+\theta, \arctan\left( \frac{r_n \sin(rt_i + \theta) + e_2}{r_n \cos(rt_i + \theta) + e_1} \right) \right)^2
    \le Cn\|e\|_2^2.
\end{equation*}

Applying the above bound with $e=(\widetilde{E}_n)_i$ results in 
\begin{equation*}
    \rm{d} \left( \hat{t}_i, rt_i+ \theta \right)^2
    \le Cn\|(\widetilde{E}_n)_i\|_2^2.
\end{equation*}
Therefore, we can obtain that 
\begin{equation*}
    \frac{1}{n} \sum_{i=1}^n \rm{d} \left( \hat{t}_i, rt_i + \theta \right)^2
    \le C\sum_{i=1}^n \|(\widetilde{E}_n)_i\|_2^2
    = C\|\widetilde{E}_n\|_F^2
    \le CR_n^2.
\end{equation*}
Taking the square root and then minimizing over the alignment parameters yields the desired bound
\begin{equation*}
    \rm{Error}(\hat{t}, t) 
    \le C \left( (1 + \alpha) \left( \tau + \frac{\sqrt{\log(n)}}{\sqrt{n} \tau^{5/4}} + \frac{1}{c_{\rm{gap}}} \frac{\rho_{h,\tau,p}}{n\tau^{3/2}} \right) +\frac{U_S}{c_{\rm{gap}}\tau} \right).
\end{equation*}
This completes the proof of Theorem \ref{Main_theo}.

\section{Implementation details for Section~\ref{sec:simulation} and Section~\ref{sec:real}}\label{appsec:simudetails}

In this section, we provide simulation and real data details for Sections \ref{sec:simulation} and \ref{sec:real}, respectively. All numerical simulations are carried out on a 13th-generation Intel Core i9-13900HX processor. The simulation packages/functions are from Python, and we use the format \texttt{Python\_Functions} to illustrate. 

In the numerical analysis, to construct the similarity matrix, we employ an adjusted kernel 
\begin{equation*}
    k^\star(x, y; h) = \exp\left\{ - \frac{\|x - y\|^2_2}{2h^2} \right\} 
\end{equation*}
instead of $k(x, y; h)$ defined in \eqref{eq:gaussian-kernel}. 
Note that $k^\star(x, y; h) = (2\pi h^2)^{-1/2}k(x, y; h)$. Hence, the similarity matrix $S^\star = (2\pi h^2)^{-1/2} S$ and $\Lza = (2\pi h^2)^{(2\alpha-1)/2} L$. It is the same with the original $L$ when $\alpha = 1/2$ and has a scaling difference for $\alpha \neq 1/2$. The leading eigenvectors are still the same. Such simplification is to avoid the floating errors in computation. 

The timestamp estimation error is evaluated as
\begin{equation*}
    \mathrm{Err}(\hat{t}, t) := \min_{r\in\{-1, 1\}, \theta\in[0, 2\pi]} \frac{1}{\sqrt{n}} \left\| [rt+\theta - \hat{t}\,]_{2\pi} \right\|_2.
\end{equation*}

The ranking error is defined as
\begin{equation*}
    \mathrm{Err}(\hat{\pi},\pi)
    = \min_{\substack{r\in\{\pm1\}\\ k\in\{0,\ldots,N-1\}}}  \left[ \frac{1}{N}\sum_{i=1}^{N} \left( \frac{d_N(\hat{\pi}(i)+k,r\pi(i))}{N} \right)^2 \right]^{1/2},
\end{equation*}
where $d_N(a,b):=\min_{j\in\mathbb{Z}}|a-b+jN|$.

\subsection{Tuning parameters in Section~\ref{subsec:experiment1}}\label{appsubsec:exp1}
The only tuning parameter in Section~\ref{subsec:experiment1} is the bandwidth $h$. For each observation $Z_i$, we record the $22$nd nearest neighbor and the corresponding distance. Hence, we have $n$ distances. The bandwidth $h$ is the median value of these distances. Such $h$ is the same for all 3 variants. 

\subsection{Selection of $\alpha$ for the extra-normalized generalized Laplacian}\label{appsubsec:alpha}

To select an empirical default for $L_{ng}$, we examine the optimal $\alpha$ across different noise levels $\sigma$ and heteroscedasticity levels $\eta$. For each simulation replicate, we select the candidate $\alpha$ that minimizes the timestamp estimation error. Table~\ref{tabN2026:alpha_selection} reports the median and the 25th and 75th percentiles of the selected values.

\begin{table}[h]
\caption{Median optimal $\alpha$ with ($25\%$ quantile, $75\%$ quantile).}
\label{tabN2026:alpha_selection}
\centering
\scriptsize
\begin{tabular}{lcccccc}
\toprule
& $\sigma = .050$ & $\sigma = .075$ & $\sigma = .100$ & $\sigma = .125$ & $\sigma = .150$ \\
\midrule
$\eta = .30$
& \makecell{.800 \space (.700, .900)}
& \makecell{.700 \space (.600, .800)}
& \makecell{.700 \space (.500, .800)}
& \makecell{.650 \space (.400, .900)}
& \makecell{.600 \space (.400, 1.000)}\\
\addlinespace[3pt]
$\eta = .45$
& \makecell{.700 \space (.600, .700)}
& \makecell{.600 \space (.500, .700)}
& \makecell{.600 \space (.500, .700)}
& \makecell{.600 \space (.500, .800)}
& \makecell{.600 \space (.400, .800)}\\
\addlinespace[3pt]
$\eta = .60$
& \makecell{.600 \space (.500, .600)}
& \makecell{.600 \space (.500, .600)}
& \makecell{.600 \space (.500, .700)}
& \makecell{.600 \space (.500, .725)}
& \makecell{.650 \space (.500, .800)}\\
\addlinespace[3pt]
$\eta = .75$
& \makecell{.600 \space (.500, .600)}
& \makecell{.600 \space (.500, .600)}
& \makecell{.600 \space (.500, .700)}
& \makecell{.600 \space (.500, .700)}
& \makecell{.600 \space (.500, .800)}\\
\bottomrule
\end{tabular}
\end{table}

When both $\sigma$ and $\eta$ are low, the selected values exhibit relatively wide interquartile ranges, indicating substantial variation in the minimizing $\alpha$ across replicates. Together with the low-error regions observed in Section~\ref{subsec:experiment1}, this suggests that precise tuning is less critical in these settings. As $\sigma$ and $\eta$ increase, the median optimal values generally concentrate around $0.6$, with narrower interquartile ranges. Although no single value is optimal in every setting, these results support $\alpha=0.6$ as an empirical default for $L_{ng}$. We therefore fix this value throughout Section~\ref{subsec:experiment2} and Section~\ref{sec:real} without further dataset-specific tuning.

\subsection{Tuning parameters in Section \ref{subsec:experiment2}}\label{appsubsec:exp2}
We choose $k$ in a data-adaptive way so that the formed network has only one connected component. Starting from $k_{\min}=8$, we increase $k$ until the symmetrized graph becomes connected, as checked by \texttt{scipy.sparse.csgraph.connected\_components}. The connected graph is constructed using \texttt{sklearn.neighbors.kneighbors\_graph} with \texttt{mode="connectivity"} and \texttt{include\_self=False}. We add a buffer of 2 neighbors, and the search is capped at $\lfloor 0.15n \rfloor$. The smallest $k$ such that the network is connected will be taken as the neighbours/perplexity in algorithms. 

For the diffusion map comparison (DM), we construct the transition matrix $P$ from the similarity matrix $S$. Let $d_i=\sum_{j=1}^n S_{ij}$ and define
\begin{equation*}
    A_{ij}=\frac{S_{ij}}{d_i^\alpha d_j^\alpha},
\qquad
\tilde d_i=\sum_{j=1}^n A_{ij},
\qquad
P_{ij}=\frac{A_{ij}}{\tilde d_i},
\end{equation*}
with $\alpha=0.5$. Thus, $P$ is the row-normalized transition matrix associated with the reweighted similarity matrix $A$. We extract the two right eigenvectors associated with the largest nontrivial eigenvalues of $P$, excluding the constant eigenvector corresponding to eigenvalue $1$, and apply our spectral seriation procedure to recover timestamps and the circular ordering.

For spectral seriation methods (baseline, DM, and $\Lng(0.6)$), we set the bandwidth $h$ as the median value of the 22nd nearest distance to each data point.

The t-SNE method is generated using \texttt{sklearn.manifold.TSNE} with 
\begin{align*}
    &\texttt{n\_components=1}, \quad 
    \texttt{perplexity=22},  \quad
    \texttt{init="pca"}, \quad 
    \texttt{learning\_rate="auto"}, \\
    &\texttt{max\_iter=1000}, \quad 
    \texttt{method="barnes\_hut"}, \quad
    \texttt{angle=0.5}, \quad
    \texttt{metrix="euclidean"}, \\
    & \texttt{random\_state=42}.
\end{align*}

The UMAP method is generated using \texttt{umap}  with 
\begin{align*}
    &\texttt{n\_components=1}, \quad
    \texttt{n\_neighbors=22}, \quad 
    \texttt{init="spectral"}, \quad
    \texttt{min\_dist=0.1}, \\
    &\texttt{metrix="euclidean"}, \quad 
    \texttt{random\_state=42}, \quad
    \texttt{n\_jobs=1}.
\end{align*}

The Isomap method is generated using \texttt{sklearn.manifold.Isomap} with 
\begin{align*}
    &\texttt{n\_components=1}, \quad
    \texttt{n\_neighbors=22}, \quad
    \texttt{metrix="euclidean"}, \quad
    \texttt{n\_jobs=1}, \\
    &\texttt{path\_method="auto"}, \quad
    \texttt{eigen\_solver="auto"}, \quad 
    \texttt{neighbors\_algorithm="auto"}.
\end{align*}

\subsection{Pink Angel data}\label{appsubsec:angel}

Each image has $144 \times 192$ pixels, where each pixel has a 3-dimensional RGB vector. Let $Z^{(R)} \in \mathbb{R}^{72 \times 27648}$ record the value in red channels, where the $i$th row represents the flattened vector from image $i$ with $144 \times 192 = 27648$ pixels. We normalize $Z^{(R)}$ so that each column has a mean of $0$ and divide it by 255 so that all values are between 0 and 1. Then we embed it into the principal space formed by the first 20 principal components. The resulting matrix is denoted as $Z^{(R)}_{pca} \in \mathbb{R}^{72 \times 20}$. We then concatenate all the three channels to form $Z_{\mathrm{pca}} = \bigl[ Z^{(R)}_{\mathrm{pca}}, Z^{(G)}_{\mathrm{pca}}, Z^{(B)}_{\mathrm{pca}} \bigr] \in \reals^{72\times 60}$. 

\begin{figure}[htb]
    \centering
    \includegraphics[width=\linewidth]{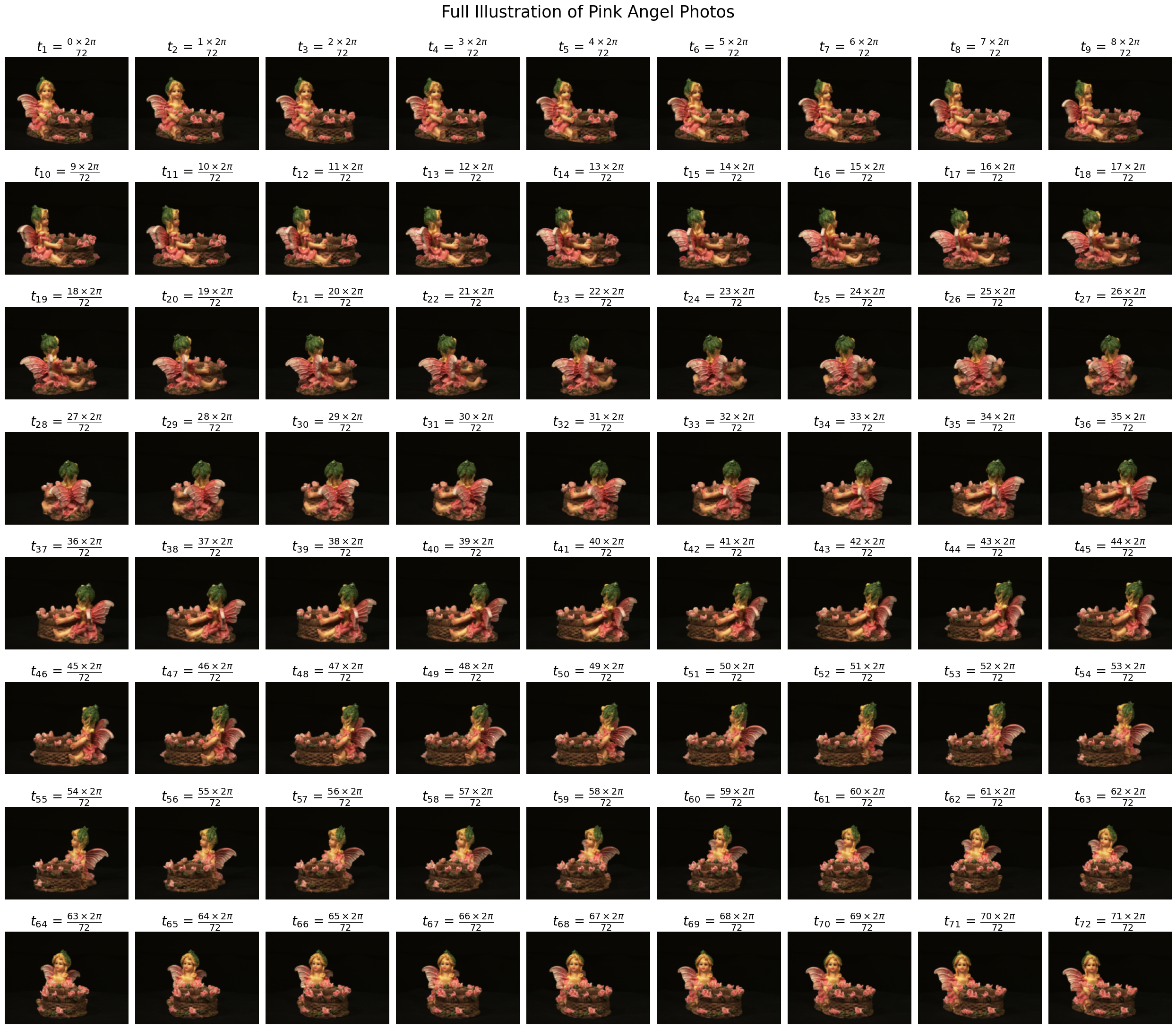}
    \caption{A full illustration of the Angel images. The viewing angle is stated above each subplot.}
    \label{figN2026:Angel_FULL}
\end{figure}

For spectral seriation methods (baseline, DM, and $\Lng(0.6)$), we set the bandwidth $h$ as the median value of the 8th nearest distance to each data point to construct the similarity matrix from input data $Z_{\mathrm{pca}}$. The comparison methods in this section use the same settings as in Section \ref{subsec:experiment2}, with \texttt{perplexity/n\_neighbors = 8}.

For completeness, we provide a full illustration of the Pink Angel data in Figure \ref{figN2026:Angel_FULL}. The dataset contains 72 images of the same object, each observed from a different viewing angle.

\subsection{FUCCI Data}\label{appsubsec:FUCCI}
\cite{FUCCI_cite} provides the filtered data, named 'GSE121265\_eset-final.rds.gz'. This dataset contains 888 cells and 11,093 expression features. We retain the 11,040 human genes whose Ensembl identifiers begin with “ENSG,” excluding spike-in controls and reporter genes. For cell $i$ and gene $g$, we transform the expression count $C_{ig}$ as
\begin{equation*}
    Z_{ig}=\log_2\left(1+\frac{10^6C_{ig}}{m_i}\right),
\end{equation*}
where $m_i$ is the molecule count supplied in the cell metadata. Reference phases are taken directly from the supplied theta field and used for evaluation.

The six cell lines—NA18511, NA18855, NA18870, NA19098, NA19101, and NA19160—contain 103, 163, 206, 182, 122, and 112 cells, respectively. We align expression profiles with cell annotations before exporting them to Python. For the pooled experiment, we concatenate the six expression matrices after verifying that their gene identifiers and column orders agree.

We reduce dimensionality separately for each of the seven experiments using a randomized SVD projection. For the gene-by-cell matrix $X=Z^\top$, we generate a standard Gaussian matrix $G\in\mathbb{R}^{n\times50}$ with random seed 0 and compute the SVD of $XG$. We retain left singular vectors whose corresponding singular values are at least $\eta$ times the largest singular value, and use $U^\top X$ as the reduced representation. We set $\eta=0.02$. This yields 49 dimensions for NA18511 and NA19101 and 50 for all other experiments. No additional centering or gene-wise standardization is applied.

All compared methods use the same reduced representation within each experiment. For the spectral methods, the Gaussian-kernel bandwidth parameter is the median positive pairwise Euclidean distance. We use $\alpha=0.6$ for $L_{\mathrm{ng}}$ and a normalization parameter of 0.5 for diffusion maps. For individual cell lines, t-SNE uses perplexity 10, while UMAP and Isomap use 10 neighbors; these values are increased to 30 for the pooled experiment.

\section{Additional Simulations under the Main Setup}\label{appsec:addsimu}

Throughout this section, we use the same setting as Section \ref{subsec:experiment1}. Consider a synthetic closed-loop manifold in $\reals^{40}$. We generate the data points as 
\begin{equation*}
    Z_i 
    = X(t_i) + s(t_i)e_i
    = Qv(t_i) + s(t_i)e_i, 
    \quad t_i \sim \mathrm{Unif}(0, 2\pi), 
    \quad 1 \leq i \leq n,
\end{equation*}
where $e_i \sim \mathcal{N}(\zero, \sigma^2I_{40})$, $v(t)=(1.5\cos(t), \sin(t), \cos(2t), \sin(2t))^\top$, $s(t) = 1 + \eta \cos(t - \pi/3)$, $X(t) = Qv(t) \in \reals^40$, and $Q^\top Q = I_r$. 

\subsection{Changing Sample Size}\label{appsubsec:sampleSize}

\begin{figure}[h]
    \centering
    \includegraphics[width=0.9\linewidth]{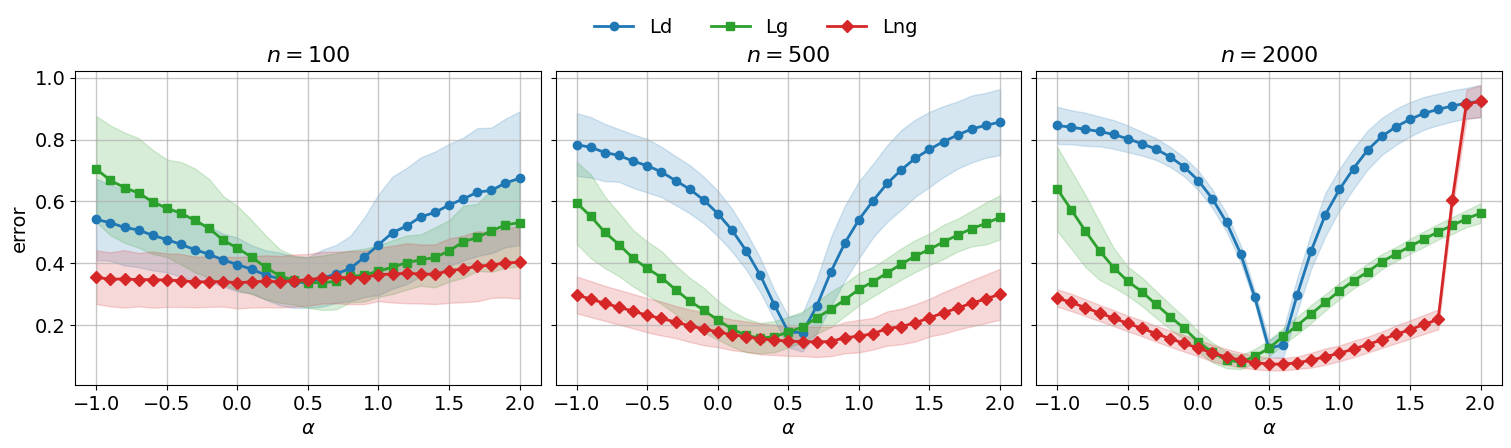}
    \caption{Mean timestamp estimation error versus $\alpha$ with $\eta=0.3$ and $\sigma=0.1$.}
    \label{figN2026:app_n1}
\end{figure}

\begin{figure}[h]
    \centering
    \includegraphics[width=0.9\linewidth]{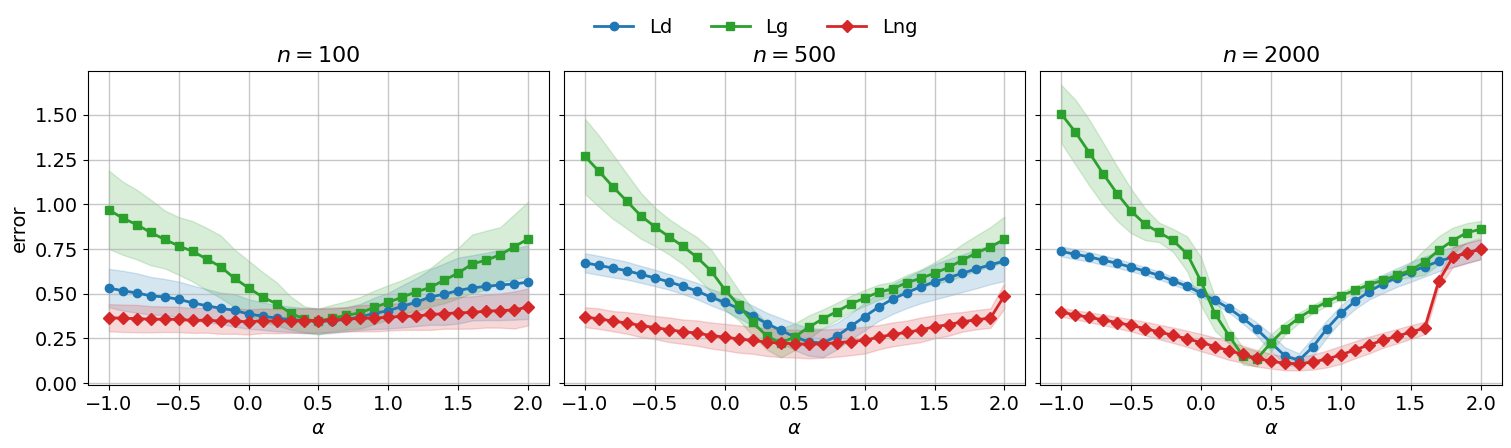}
    \caption{Mean timestamp estimation error versus $\alpha$ with $\eta=0.3$ and $\sigma=0.15$.}
    \label{figN2026:app_n2}
\end{figure}

\begin{figure}[h]
    \centering
    \includegraphics[width=0.9\linewidth]{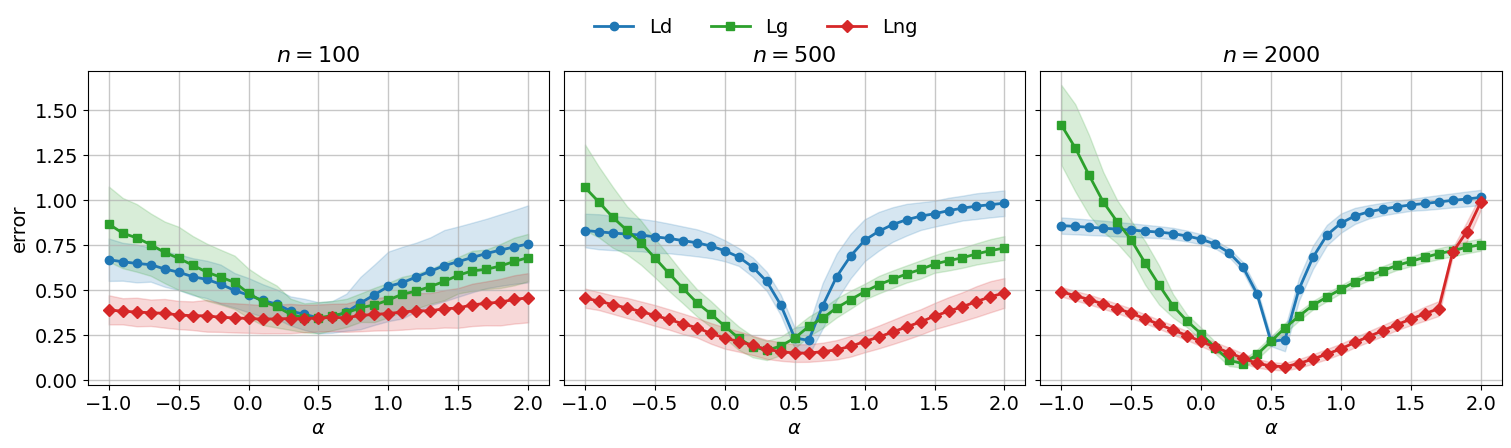}
    \caption{Mean timestamp estimation error versus $\alpha$ with $\eta=0.6$ and $\sigma=0.1$.}
    \label{figN2026:app_n3}
\end{figure}

We examine the effect of sample size by varying $n\in\{100,500,2000\}$ under three settings: $(\eta,\sigma)=(0.3,0.1)$, $(0.3,0.15)$, and $(0.6,0.1)$, as shown in Figures~\ref{figN2026:app_n1}-~\ref{figN2026:app_n3}, respectively. Across these settings, the timestamp estimation error of $\Lng$ generally decreases with increasing sample size for $\alpha\in[0,1]$, consistent with the asymptotic results in Section~\ref{sec:theory}. As the sample size increases, the differences in sensitivity to $\alpha$ become more pronounced. While $\Ld$ exhibits a sharp error minimum and $\Ld$ can suffer large errors for negative $\alpha$, $\Lng$ maintains low estimation error over a broader range of $\alpha$. This advantage persists under higher noise and heterogeneity, with the latter particularly amplifying the sensitivity of $\Lg$ to negative $\alpha$. However, for sufficiently large $\alpha$, $\Lng$ can still exhibit a sharp increase in error even at $n=2000$. These results further demonstrate the robustness of $\Lng$ to parameter tuning, while highlighting the importance of choosing $\alpha$ within an appropriate range.

\subsection{Bandwidth Sensitivity}\label{appsubsec:bandwdith}

We examine the sensitivity of the three generalized Laplacians to bandwidth choices around the values selected by the rule described in Appendix~\ref{appsubsec:exp1}. The reference bandwidths are $0.9488$, $1.349$, and $0.9428$ for $(\eta,\sigma)=(0.3,0.1)$, $(0.3,0.15)$, and $(0.6,0.1)$, respectively. Tables~\ref{tabN20267:bandwidth1}-~\ref{tabN20267:bandwidth3} report the results for bandwidths around these values. The estimation error varies with the bandwidth, and the sensitivity depends on both the Laplacian variant and $\alpha$. Nevertheless, for $\alpha\in[0,1]$, $\Lng$ generally achieves lower estimation errors than $\Ld$ and $\Lg$ across the tested bandwidths, although its error can also vary substantially. For larger $\alpha$, this advantage may no longer hold. These results suggest that the improved estimation accuracy of $\Lng$ is not restricted to a particular bandwidth choice, while appropriate bandwidth selection remains important.

\begin{table}[h]
\centering
\caption{Mean timestamp estimation errors (standard deviations) under different
bandwidths and values of $\alpha$, with $\eta=0.3$ and $\sigma=0.1$.}
\label{tabN20267:bandwidth1}
\scriptsize
\setlength{\tabcolsep}{4pt}
\begin{tabular}{clccccc}
\toprule
& & \multicolumn{5}{c}{Bandwidth} \\
\cmidrule(lr){3-7}
$\alpha$ & Method & .75 & .90 & 1.05 & 1.20 & 1.35 \\
\midrule
$-0.5$
& $\Ld$ & 1.0187 (.0950) & .8511 (.1126) & .6671 (.1261) & .5108 (.1224) & .4127 (.1112) \\
& $\Lg$ & .3595 (.0356) & .3443 (.0455) & .3899 (.0680) & .5101 (.1474) & .6493 (.1773) \\
& $\Lng$ & \textbf{.2423} (.0489) & \textbf{.2278} (.0568) & \textbf{.2278} (.0648) & \textbf{.2363} (.0732) & \textbf{.2534} (.0895) \\
\midrule
$0$
& $\Ld$ 
& .8412 (.0868) & .6558 (.0991) & .4782 (.1075) & .3717 (.0979) & .3195 (.0947) \\
& $\Lg$ 
& .1476 (.0210) & .1546 (.0299) & .1926 (.0439) & .2664 (.0980) & .3998 (.1797) \\
& $\Lng$ 
& \textbf{.1484} (.0336) & \textbf{.1528} (.0417) & \textbf{.1694} (.0521) & \textbf{.1937} (.0661) & \textbf{.2365} (.1206) \\
\midrule
$0.5$
& $\Ld$ 
& .1484 (.0336) & .1528 (.0417) & .1694 (.0521) & .1937 (.0661) & .2365 (.1206) \\
& $\Lg$ 
& .1484 (.0336) & .1528 (.0417) & .1694 (.0521) & .1937 (.0661) & .2365 (.1206) \\
& $\Lng$ 
& \textbf{.0775} (.0170) & \textbf{.0963} (.0255) & \textbf{.1252} (.0397) & \textbf{.1626} (.0620) & \textbf{.2221} (.1428) \\
\midrule
$1$
& $\Ld$ 
& .8633 (.0512) & .5999 (.0730) & .3847 (.0769) & .2847 (.1002) & .3022 (.1822) \\
& $\Lg$ 
& .3367 (.0631) & .3240 (.0736) & .3316 (.0826) & .3398 (.0886) & .3419 (.0921) \\
& $\Lng$ 
& \textbf{.1189} (.0107) & \textbf{.1101} (.0204) & \textbf{.1249} (.0330) & \textbf{.1643} (.0695) & \textbf{.2414} (.1625) \\
\midrule
$1.5$
& $\Ld$ 
& 1.0736 (.0638) & .8414 (.0751) & .6401 (.0988) & .5155 (.1147) & .4823 (.1600) \\
& $\Lg$ 
& .4966 (.0838) & .4695 (.0950) & .4639 (.1034) & .4552 (.1068) & .4436 (.1068) \\
& $\Lng$
& \textbf{.2164} (.0147) & \textbf{.1827} (.0193) & \textbf{.1730} (.0352) & \textbf{.1942} (.0843) & \textbf{.2761} (.1857) \\
\bottomrule
\end{tabular}
\end{table}

\begin{table}[h]
\centering
\caption{Mean timestamp estimation errors (standard deviations) under different
bandwidths and values of $\alpha$, with $\eta=0.3$ and $\sigma=0.15$.}
\label{tabN20267:bandwidth2}
\scriptsize
\setlength{\tabcolsep}{4pt}
\begin{tabular}{clccccc}
\toprule
& & \multicolumn{5}{c}{Bandwidth} \\
\cmidrule(lr){3-7}
$\alpha$ & Method & 1.05 & 1.20 & 1.35 & 1.50 & 1.65 \\
\midrule
$-0.5$
& $\Ld$ 
& .8980 (.0680) & .7540 (.0737) & .6321 (.0774) & .5449 (.0815) & .4872 (.0870) \\
& $\Lg$ 
& .8344 (.1145) & .8682 (.1182) & .8910 (.1176) & .9142 (.1103) & .9247 (.1078) \\
& $\Lng$ 
& \textbf{.3769} (.0673) & \textbf{.3672} (.0848) & \textbf{.3719} (.1163) & \textbf{.3857} (.1562) & \textbf{.4029} (.1849) \\
\midrule
$0$
& $\Ld$ 
& .7229 (.0686) & .5870 (.0760) & .4959 (.0856) & .4448 (.0986) & .4215 (.1139) \\
& $\Lg$ 
& .3619 (.0584) & .4620 (.1210) & .5569 (.1449) & .6412 (.1557) & .7032 (.1584) \\
& $\Lng$ 
& \textbf{.2551} (.0603) & \textbf{.2746} (.0819) & \textbf{.3215} (.1479) & \textbf{.3660} (.2015) & \textbf{.4052} (.2307) \\
\midrule
$0.5$
& $\Ld$ 
& .2551 (.0603) & .2746 (.0819) & .3215 (01479) & .3660 (.2015) & \textbf{.4052} (.2307) \\
& $\Lg$ 
& .2551 (.0603) & .2746 (.0819) & .3215 (.1479) & .3660 (.2015) & \textbf{.4052} (.2307) \\
& $\Lng$ 
& \textbf{.1469} (.0418) & \textbf{.2011} (.1221) & \textbf{.2724} (.1723) & \textbf{.3595} (.2504) & .4334 (.2954) \\
\midrule
$1$
& $\Ld$ 
& .6469 (.0518) & .4757 (.0669) & .4198 (.1490) & .4528 (.2238) & .5212 (.2909) \\
& $\Lg$ 
& .5378 (.0653) & .5252 (.0687) & .5116 (.0695) & .4985 (.0692) & .4872 (.0686) \\
& $\Lng$ 
& \textbf{.1888} (.0413) & \textbf{.2129} (.0835) & \textbf{.2956} (.1862) & \textbf{.3962} (.2710) & \textbf{.4803} (.3068) \\
\midrule
$1.5$
& $\Ld$ 
& .9144 (.0482) & .7513 (.0579) & .6357 (.0736) & .5932 (.1180) & .6217 (.1859) \\
& $\Lg$ 
& .6949 (.0737) & .6653 (.0727) & .6415 (.0777) & .6480 (.0872) & .6788 (.0923) \\
& $\Lng$ 
& \textbf{.3257} (.0386) & \textbf{.3146} (.0801) & \textbf{.3697} (.1797) & \textbf{.4593} (.2522) & \textbf{.5377} (.2916) \\
\bottomrule
\end{tabular}
\end{table}

\begin{table}[h]
\centering
\caption{Mean timestamp estimation errors (standard deviations) under different
bandwidths and values of $\alpha$, with $\eta=0.6$ and $\sigma=0.1$.}
\label{tabN20267:bandwidth3}
\scriptsize
\setlength{\tabcolsep}{4pt}
\begin{tabular}{clccccc}
\toprule
& & \multicolumn{5}{c}{Bandwidth} \\
\cmidrule(lr){3-7}
$\alpha$ & Method & .75 & .90 & 1.05 & 1.20 & 1.35 \\
\midrule
$-0.5$
& $\Ld$ 
& 1.1412 (.0663) & 1.0062 (.0684) & .8582 (.0813) & .7123 (.0836) & .5929 (.0856) \\
& $\Lg$ 
& .6905 (.0821) & .6577 (.0962) & .7001 (.1072) & .7529 (.1094) & .8025 (.1097) \\
& $\Lng$ 
& \textbf{.4079} (.0464) & \textbf{.3673} (.0556) & \textbf{.3441} (.0656) & \textbf{.3331} (.0777) & \textbf{.3356} (.0981) \\
\midrule
$0$
& $\Ld$ 
& .9894 (.0623) & .8348 (.0693) & .6797 (.0774) & .5469 (.0836) & .4598 (.0910) \\
& $\Lg$ 
& .2567 (.0204) & .2612 (.0281) & .3059 (.0363) & .3726 (.0487) & .4591 (.0936) \\
& $\Lng$ 
& \textbf{.2345} (.0359) & \textbf{.2257} (.0457) & \textbf{.2310} (.0575) & \textbf{.2468} (.0742) & \textbf{.2791} (.1166) \\
\midrule
$0.5$
& $\Ld$ 
& .2345 (.0359) & .2257 (.0457) & .2310 (.0575) & .2468 (.0742) & .2791 (.1166) \\
& $\Lg$ 
& .2345 (.0359) & .2257 (.0457) & .2310 (.0575) & .2468 (.0742) & .2791 (.1166) \\
& $\Lng$ 
& \textbf{.0799} (.0167) & \textbf{.1003} (.0261) & \textbf{.1309} (.0416) & \textbf{.1697} (.0640) & \textbf{.2322} (.1494) \\
\midrule
$1$
& $\Ld$ 
& 1.0743 (.0323) & .8428 (.0364) & .6038 (.0512) & .4288 (.0627) & .3657 (.1242) \\
& $\Lg$ 
& .5458 (.0527) & .5093 (.0602) & .4982 (.0681) & .4903 (.0741) & .4792 (.0751) \\
& $\Lng$ 
& \textbf{.2129} (.0133) & \textbf{.1742} (.0247) & \textbf{.1629} (.0400) & \textbf{.1849} (.0700) & \textbf{.2499} (.1508) \\
\midrule
$1.5$
& $\Ld$ 
& 1.1875 (.0875) & 1.0695 (.0375) & .8807 (.0463) & .7081 (.0580) & .5927 (.0729) \\
& $\Lg$ 
& .7007 (.0668) & .6686 (.0721) & .6493 (.0766) & .6266 (.0777) & .6032 (.0781) \\
& $\Lng$ 
& \textbf{.4042} (.0163) & \textbf{.3331} (.0218) & \textbf{.2879} (.0355) & \textbf{.2746} (.0684) & \textbf{.3197} (.1685) \\
\bottomrule
\end{tabular}
\end{table}

\newpage

\subsection{Different Similarity Constructions}\label{appsubsec:DiffSimConst}

We examine how the construction of the similarity matrix affects phase recovery by comparing the Gaussian kernel with two sparse alternatives: RBF-weighted KNN and unweighted KNN.

For the Gaussian construction, we use $S_{ij}=k(Z_i,Z_j;h)$ as defined in \eqref{eq:gaussian-kernel}. For the two KNN constructions, let $\mathcal{N}_k(i)$ denote the indices of the $k$ nearest neighbors of $Z_i$ in Euclidean distance, excluding $i$. We define the symmetric neighborhood indicator
\begin{equation*}
    M_{ij}
    = \one \bigl\{ j\in\mathcal{N}_k(i) \ \text{or}\ i\in\mathcal{N}_k(j) \bigr\},
    \qquad i\neq j,
\end{equation*}
with $M_{ii}=0$. Thus, two observations are connected whenever either belongs to the $k$-nearest-neighbor set of the other. For RBF-weighted KNN, we retain the Gaussian kernel values on the selected edges:
\begin{equation*}
    S_{ij}^{\mathrm{KNN}}
    = M_{ij}\,k(Z_i,Z_j;h).
\end{equation*}
For unweighted KNN, all selected edges receive unit weight:
\begin{equation*}
    S_{ij}^{\mathrm{unKNN}}
    = M_{ij}.
\end{equation*}
Both KNN matrices therefore have zero diagonal and use the same neighborhood structure. Weighted KNN retains distance-dependent edge weights, whereas unweighted KNN uses only local connectivity. We apply the three generalized Laplacian constructions and the same phase-recovery procedure to each similarity matrix.

Table~\ref{tabN2026:Similarity} reports the mean timestamp estimation errors and standard deviations. Across all three similarity constructions, $\Lng$ demonstrates favorable performance in terms of both estimation accuracy and robustness to $\alpha$. For the Gaussian kernel, $\Lng$ achieves the lowest mean error at every tested $\alpha$. For RBF-weighted KNN, $\Lng$ achieves lower errors for $\alpha \geq 0.5$. While for unweighted KNN, its error remains nearly constant across the tested values of $\alpha$. Although $\Lg$ can achieve lower errors at certain values of $\alpha$ under KNN constructions, $\Lng$ exhibits the smallest error variation and lowest maximum mean error across $\alpha$ under all three constructions. These results demonstrate that extra normalization provides a favorable balance between estimation accuracy and robustness to the choice of $\alpha$, regardless of the similarity construction.

\begin{table}[h]
\centering
\caption{Timestamp estimation errors (with standard deviation).}
\label{tabN2026:Similarity}
\scriptsize
\begin{tabular}{c|*{3}{c}|*{3}{c}|*{3}{c}}
\toprule
& \multicolumn{3}{c|}{Gaussian Kernel}
& \multicolumn{3}{c|}{KNN}
& \multicolumn{3}{c}{unweighted KNN} \\
\midrule
$\alpha$
& $\Lda$ & $\Lga$ & $\Lnga$ 
& $\Lda$ & $\Lga$ & $\Lnga$
& $\Lda$ & $\Lga$ & $\Lnga$ \\
\midrule
$-0.5$
& \makecell{.7873 \\ (.1347)}
& \makecell{.3529 \\ (.0482)}
& \makecell{\textbf{.2268} \\ (.0592)}
& \makecell{1.0178 \\ (.1149)}
& \makecell{.0968 \\ (.0142)}
& \makecell{\textbf{.1852} \\ (.0368)}
& \makecell{.8094 \\ (.1098)}
& \makecell{.1274 \\ (.0220)}
& \makecell{\textbf{.1071} \\ (.0222)} \\
\addlinespace[3pt]
$0.0$
& \makecell{.5899 \\ (.1233)}
& \makecell{.1642 \\ (.0339)}
& \makecell{\textbf{.1573} \\ (.0439)}
& \makecell{.9782 \\ (.1075)}
& \makecell{\textbf{.0836} \\ (.0136)}
& \makecell{.1450 \\ (.0299)}
& \makecell{.7022 \\ (.1198)}
& \makecell{.1170 \\ (.0229)}
& \makecell{\textbf{.1060} \\ (.0231)} \\
\addlinespace[3pt]
$0.5$
& \makecell{.1573 \\ (.0439)}
& \makecell{.1573 \\ (.0439)}
& \makecell{\textbf{.1047} \\ (.0292)}
& \makecell{.1450 \\ (.0299)}
& \makecell{.1450 \\ (.0299)}
& \makecell{\textbf{.1063} \\ (.0224)}
& \makecell{\textbf{.1060} \\ (.0231)}
& \makecell{\textbf{.1060} \\ (.0231)}
& \makecell{\textbf{.1077} \\ (.0237)} \\
\addlinespace[3pt]
$1.0$
& \makecell{.5211 \\ (.1009)}
& \makecell{.3247 \\ (.0762)}
& \makecell{\textbf{.1129} \\ (.0249)}
& \makecell{1.0895 \\ (.0667)}
& \makecell{.2144 \\ (.0439)}
& \makecell{\textbf{.0760} \\ (.0134)}
& \makecell{.9248 \\ (.1171)}
& \makecell{\textbf{.0973} \\ (.0218)}
& \makecell{.1115 \\ (.0238)} \\
\addlinespace[3pt]
$1.5$
& \makecell{.7710 \\ (.1054)}
& \makecell{.4657 \\ (.0992)}
& \makecell{\textbf{.1782} \\ (.0241)}
& \makecell{1.1746 \\ (.0930)}
& \makecell{.2760 \\ (.0573)}
& \makecell{\textbf{.0693} \\ (.0096)}
& \makecell{1.0366\\ (.1299)}
& \makecell{\textbf{.0966} \\ (.0212)}
& \makecell{.1166 \\ (.0233)} \\
\bottomrule
\end{tabular}
\end{table}

\subsection{Rotating Angle Photos using Tangent Distance}\label{appsubsec:angle_TD}

We revisit the rotating angle image experiment using tangent distance to construct the similarity matrix. This experiment examines whether accounting for small translations and changes in scale improves recovery of the underlying rotation angles. We use the preprocessing procedure described in Appendix~\ref{appsubsec:alpha} and construct a Gaussian affinity from the two-sided tangent distance~\citep{Simard2000}.

For each preprocessed image, we construct tangent directions corresponding to horizontal translation, vertical translation, and isotropic scaling about the image center. Let $f$ denote the concatenated PCA feature map. The horizontal translation tangent is approximated by
\begin{equation*}
    b_{i,x}
    = \frac{f(T_{x,\delta}I_i)-f(T_{x,-\delta}I_i)}{2\delta},
    \qquad \delta=0.5,
\end{equation*}
where $T_{x,\delta}$ translates an image horizontally by $\delta$ pixels. The vertical translation tangent $b_{i,y}$ is computed analogously. For scaling, we use
\begin{equation*}
    b_{i,s}
    = \frac{f(T_{s,1+\varepsilon}I_i)-f(T_{s,1-\varepsilon}I_i)}{2\varepsilon},
    \qquad \varepsilon=0.01,
\end{equation*}
where $T_{s,c}$ scales an image by a factor of $c$ about its center. All transformations use bilinear interpolation, with constant background values estimated from the median boundary intensity of each color channel. We orthonormalize $[b_{i,x},b_{i,y},b_{i,s}]$ using SVD, retaining singular directions whose singular values exceed $10^{-8}$ times the largest singular value. This gives a tangent basis $B_i\in\mathbb{R}^{60\times r_i}$, where $r_i\leq3$. Rotation is not included among the tangent transformations.

For each pair of images, we compute the squared two-sided tangent distance
\begin{equation*}
    d_{\mathrm{tan}}^2(i,j)
= \min_{a\in\reals^{r_i},\,b\in\reals^{r_j}}
\left\|z_i+B_i a-z_j-B_j b\right\|_2^2.
\end{equation*}
This least-squares residual measures the feature discrepancy remaining after accounting for the local translation and scaling directions of both images. We then construct the Gaussian similarity matrix
\begin{equation*}
    S_{ij}
    = \exp \left(-\frac{d_{\mathrm{tan}}^2(i,j)}{2h^2}\right).
\end{equation*}

Let $r_{i,k}$ denote the tangent distance from image $i$ to its $k$th nearest neighbor, excluding itself. We select the global bandwidth as $h=\left(\mathrm{median}_{i\in[n]}r_{i,k}\right)^{1/2}$, $k=\min\{20,n-1\}$. The resulting similarity matrix is symmetric, has $S_{ii}=1$, and retains all pairwise similarities without nearest-neighbor truncation.

Table~\ref{tabN2026:Angel_TD} compares the results obtained using Euclidean and tangent distances. Tangent distance substantially reduces the ordering errors of t-SNE, UMAP, and Isomap, from $0.1857$, $0.1873$, and $0.1927$ to $0.0343$, $0.0249$, and $0.0385$, respectively. The three spectral methods recover the circular ordering exactly under both constructions. However, their timestamp estimation errors increase when tangent distance is used. Among these methods, $L_{ng}(0.6)$ achieves the lowest timestamp estimation error under both constructions. Thus, accounting for translation and scale variation improves ordering for the manifold-learning comparisons, but does not improve phase estimation in this experiment.

\begin{table}[h]
\centering
\caption{Ordering and timestamp estimation errors using Euclidean and tangent distances.}
\label{tabN2026:Angel_TD}
\scriptsize
\begin{tabular}{c|*{2}{c}|*{2}{c}}
\toprule
& \multicolumn{2}{c|}{Euclidean distances}
& \multicolumn{2}{c}{Tangent distances} \\
\midrule
Method
& $\mathrm{Error}(\hat{\pi}, \pi)$ & $\mathrm{Error}(\hat{t}, t)$ 
& $\mathrm{Error}(\hat{\pi}, \pi)$ & $\mathrm{Error}(\hat{t}, t)$ \\
\midrule
t-SNE
& .1857 
& /
& .0343
& / \\
\addlinespace[3pt]
UMAP
& .1873
& /
& .0249
& / \\
\addlinespace[3pt]
Isomap
& .1927 
& /
& .0385
& / \\
\addlinespace[3pt]
Baseline
& \textbf{.0}
& .1339
& \textbf{.0}
& .1947 \\
\addlinespace[3pt]
Diffusion Maop
& \textbf{.0}
& .1042
& \textbf{.0}
& .1258 \\
\addlinespace[3pt]
$\Lng(0.6)$
& \textbf{.0}
& \textbf{.1004}
& \textbf{.0}
& \textbf{.1143} \\
\bottomrule
\end{tabular}
\end{table}

\section{Non-uniform Sampling of Latent Phases}\label{appsec:nonUnif}

We further investigate the performance of the three Laplacian constructions under non-uniform sampling of latent phases. While the main analysis assumes uniformly distributed phases, we consider a more general sampling setting to examine whether accurate recovery can still be achieved.

We generate observations from a synthetic closed-loop manifold in $\reals^{40}$ as
\begin{equation}
    Z_i = X(t_i) + e_i = Qv(t_i) + e_i,
    \qquad i \in [n],
\end{equation}
where $v(t) = (1.5\cos t,\, 1.5\sin t,\, \cos 2t,\, \sin 2t)^\top$, $Q \in \reals^{40 \times 4}$ satisfies $Q^\top Q=I_4$, and $e_i \sim \mathcal{N}(0,\sigma^2 I_{40})$. Unlike the uniform sampling in the main experiments, we generate the latent phases according to
\begin{equation*}
    t_i=2 \pi u_i,
    \qquad
    u_i \overset{\mathrm{i.i.d.}}{\sim} \mathrm{Beta}(a,b).
\end{equation*}
The trajectory $X(t)$ has constant speed, so the non-uniform sampling density is determined by the Beta distribution.

We fix $n=500$ and $\sigma=0.1$, and consider $a,b \in \{1, 2, 3, 5\}$. We compare the three Laplacian constructions $\Ld$, $\Lg$, and $\Lng$ over $\alpha \in \{-1.00, -0.75, \ldots, 0, 0.25,\ldots,2\}$, using both timestamp estimation error and circular ordering error.

\begin{figure}[h]
    \centering
    \includegraphics[width=\linewidth]{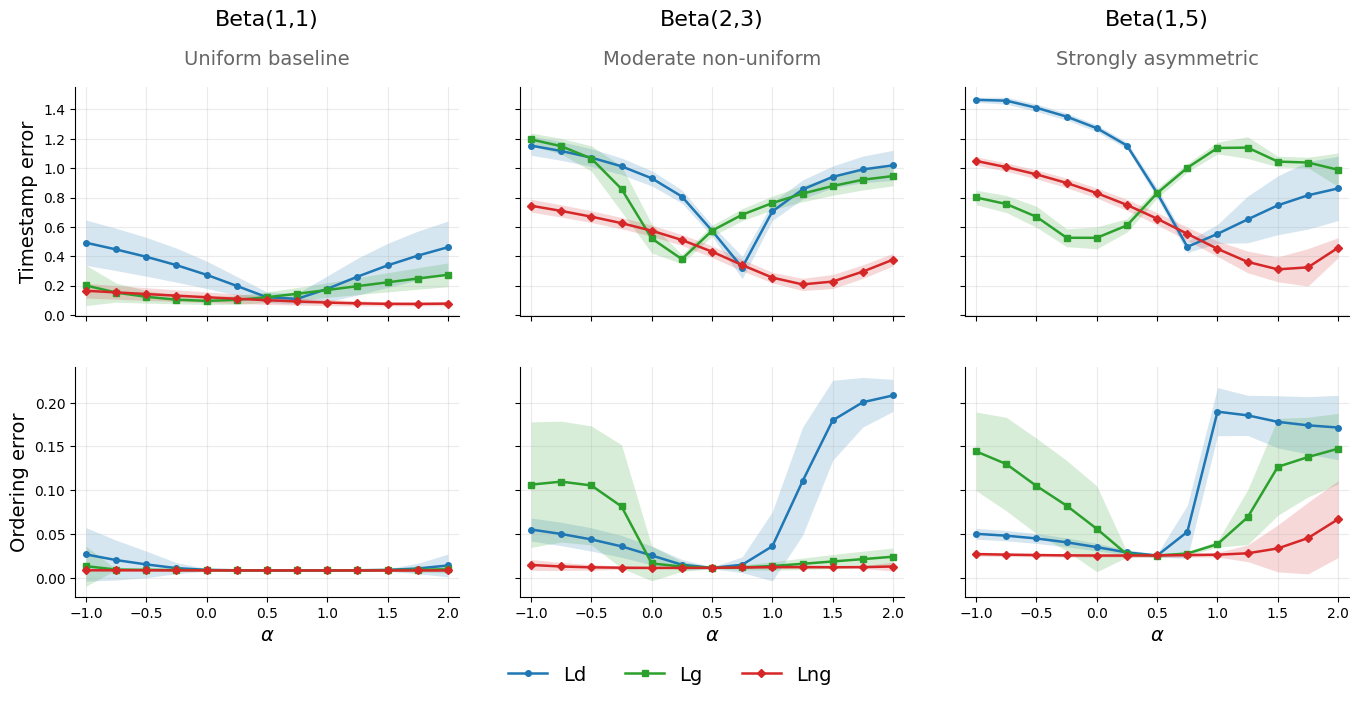}
    \caption{Estimation error under non-uniform sampling. The three columns correspond to $\mathrm{Beta}(1,1)$, $\mathrm{Beta}(2,3)$, and $\mathrm{Beta}(1,5)$, respectively. The top and bottom rows report timestamp estimation error and circular ordering error as functions of $\alpha$, respectively. Solid lines represent the mean errors, and shaded regions indicate one standard deviation across $100$ repetitions.}
    \label{figN2026:app_NonUniform}
\end{figure}

Figure~\ref{figN2026:app_NonUniform} shows the performance of the three Laplacian constructions under different Beta distributions. U. For circularnder non-uniform sampling, the timestamp estimation errors become more sensitive to the choice of $\alpha$. However, $\Lng$ still achieves lower timestamp estimation errors than the other two variants with a suitable choice of $\alpha$, even under strongly asymmetric sampling ordering, $\Lng$ maintains relatively low errors over a wide range of $\alpha$.

To further compare the recovery accuracy of the three constructions, we consider the best achievable timestamp estimation error
\begin{equation}
    E_t^\star(a,b;L)
    = \min_{\alpha \in \mathcal{A}}
    E_t(a,b;L,\alpha),
\end{equation}
where $\mathcal{A}=\{-1,-0.75,\ldots,2\}$, $L\in\{\Ld,\Lg,\Lng\}$, and $E_t(a,b;L,\alpha)$ denotes the mean timestamp estimation error across repetitions. We denote the corresponding optimal normalization parameter by $\alpha_t^\star$.

Table~\ref{tabN2026:app_nonuniform} reports the best achievable timestamp estimation errors, their standard deviations, and the optimal values of $\alpha$. We report only $a\leq b$, since interchanging $a$ and $b$ corresponds to a reflection of the latent phases under the current simulation setting. Across all ten Beta settings, $\Lng$ achieves the lowest mean estimation error, with $\alpha_t^\star$ ranging from $1.25$ to $1.75$. These results suggest that $\Lng$ remains effective under non-uniform sampling with an appropriate choice of $\alpha$. This experiment also provides empirical evidence beyond the uniform sampling setting covered by our theoretical analysis.

\begin{table}[htb]
\centering
\caption{Best achievable timestamp estimation errors under non-uniform sampling.}
\label{tabN2026:app_nonuniform}
\scriptsize

\begin{tabular}{l|cc|cc|cc}
\toprule
& \multicolumn{2}{c|}{$L_d$}
& \multicolumn{2}{c|}{$L_g$}
& \multicolumn{2}{c}{$L_{ng}$} \\
\cmidrule(lr){2-3}
\cmidrule(lr){4-5}
\cmidrule(lr){6-7}
Beta $(a,b)$
& $E_t^\star$ & $\alpha_t^\star$
& $E_t^\star$ & $\alpha_t^\star$
& $E_t^\star$ & $\alpha_t^\star$ \\
\midrule
$(1,1)$
& 0.1093 (0.0370) & 0.75
& 0.0958 (0.0254) & 0
& \textbf{0.0749} (0.0136) & 1.75 \\
\addlinespace[3pt]
$(1,2)$
& 0.3289 (0.0464) & 0.75
& 0.3152 (0.0516) & 0
& \textbf{0.1852} (0.0404) & 1.25 \\
\addlinespace[3pt]
$(1,3)$
& 0.4854 (0.0233) & 0.75
& 0.4937 (0.1385) & -0.25
& \textbf{0.2739} (0.0493) & 1.50 \\
\addlinespace[3pt]
$(1,5)$
& 0.4637 (0.0385) & 0.75
& 0.5253 (0.0747) & 0
& \textbf{0.3105} (0.0847) & 1.50 \\
\addlinespace[3pt]
$(2,2)$
& 0.3956 (0.0352) & 0.50
& 0.2507 (0.0332) & 0.25
& \textbf{0.1471} (0.0413) & 1.25 \\
\addlinespace[3pt]
$(2,3)$
& 0.3229 (0.0760) & 0.75
& 0.3801 (0.0307) & 0.25
& \textbf{0.2079} (0.0415) & 1.25 \\
\addlinespace[3pt]
$(2,5)$
& 0.4368 (0.0467) & 0.75
& 0.6140 (0.0600) & 0.25
& \textbf{0.2937} (0.0711) & 1.75 \\
\addlinespace[3pt]
$(3,3)$
& 0.4719 (0.0968) & 0.75
& 0.3731 (0.0448) & 0.25
& \textbf{0.2006} (0.0392) & 1.25 \\
\addlinespace[3pt]
$(3,5)$
& 0.3099 (0.0781) & 0.75
& 0.5937 (0.0804) & 0.25
& \textbf{0.2694} (0.0477) & 1.50 \\
\addlinespace[3pt]
$(5,5)$
& 0.3310 (0.1191) & 0.75
& 0.5488 (0.0802) & 0.25
& \textbf{0.2659} (0.0418) & 1.50 \\
\bottomrule
\end{tabular}
\end{table}

\end{document}